\documentclass{article}
\usepackage{amsmath}
\usepackage{tabularx}
\usepackage{tikz}
\usetikzlibrary{arrows, decorations.markings,
  decorations.pathmorphing, shapes, fit, backgrounds, patterns,
  calc,petri,snakes}
\definecolor{colorS}{RGB}{0,154,128}
\definecolor{colorI}{RGB}{255,32,0}
\definecolor{colorI2}{RGB}{230,154,0}
\definecolor{colorR}{RGB}{205,10,179}

\tikzstyle{box} = [draw, minimum size=3em, text centered]
\tikzstyle{bigbox}=[draw, inner sep=20pt,label={[align=right,shift={(-1.5ex,3ex)}]south east:\llap{#1}}]

\tikzstyle{decay} = [draw, ->, decorate, decoration = {snake,
  amplitude = 0.4mm, segment length=2mm, post length = 2mm, pre length
= 1mm}]

\usepackage{fullpage}

\newcommand{\order}{\mathcal{O}}
\newcommand{\diff}[2]{\frac{\mathrm{d} #1}{\mathrm{d} #2}}

\newcommand{\ave}[1]{\left \langle #1 \right \rangle}

\newcommand\Ro[1][\relax]{\ifx\relax#1 \ensuremath{\mathcal{R}_0}
  \else \ensuremath{\mathcal{R}_{0,#1}} \fi}

\title{Mathematical models of SIR disease spread with combined non-sexual and sexual transmission routes}
\author{Joel C Miller}
\begin{document}
\maketitle
\begin{abstract}
The emergence of diseases such as Zika and Ebola has highlighted the need to understand the role of sexual transmission in the spread of diseases with a primarily non-sexual transmission route.  In this paper we develop a number of low-dimensional models which are appropriate for a range of assumptions for how a disease will spread if it has sexual transmission through a sexual contact network combined with some other transmission mechanism, such as direct contact or vectors.  The equations derived provide exact predictions for the dynamics of the corresponding simulations in the large population limit.
\end{abstract}

\section{Introduction}
Many sexually transmitted diseases are known to also spread through other mechanisms, typically blood transfusion or sharing of needles. Conversely, the recently emerging diseases of Ebola~\cite{christie2015possible,mate2015molecular} and Zika~\cite{foy2011probable,musso2015potential} demonstrate that some diseases which spread primarily through other means can also have a sexual component to their spread.  

Zika is a mosquito-borne virus which can cause birth defects if a pregnant woman is infected~\cite{ventura2016zika,oliveira2016zika}.  Although more is being learned, it appears that Zika causes self-limiting infections and infected individuals appear to recover with immunity.  

Ebola is a directly transmitted disease which causes extreme morbidity and mortality~\cite{baron1983ebola,emond1977case,heymann1980ebola}.  It is spread through direct contact with bodily fluids from an infected individual.  Individuals who survive appear to gain immunity.

For both Zika and Ebola there is evidence that viable virus persists in semen long after symptoms have resolved.  So the infectious period through sexual contacts may be longer lasting than the infectious period through standard interactions.

To better predict the spread of a disease in a population, we would need to be able to capture these different modes of transmission into a model.  Unfortunately, many of the existing mathematical methods to study disease spread through a network involve a very large number of equations~\cite{EoNbook}, but in the case of susceptible-infected-recovered diseases (such as Ebola and Zika), a low-dimensional model exists.  However its structure is very different from the usual models used for other transmission mechanisms.  Consequently it is not immediately obvious that we can combine the multiple transmission mechanisms into a single mathematical model.

In this paper we begin by revisiting established models for a susceptible-infected-recovered disease spreading through mass action mixing and through a sexual contact network.   We then explore a number of models of the spread of an SIR disease through a sexual network combined with some other transmission mode.   In all of these, we assume that the transmissions through the sexual network have a longer duration than through the other mechanism.   We begin with different assumptions about how the other transmission mechanism might behave (mass action, vector borne, or another network) combined with a static random sexual network and then we use a simple mass action model for the other transmission mechanism and make different assumptions about how the sexual network is structured (changing partnerships or preferential mixing).  

\section{The models}

Throughout we will assume that $S$, $I$, and $R$ (and any subdivisions of these classes) represent the proportion of the population that are in the susceptible, infected, or recovered state.  We assume the outbreak is initialized with a fraction $\rho=0.05$ of the population chosen uniformly at random to be infected at time $t=0$.

\subsection{The standard models}
We begin by briefly reviewing the mass action SIR model and the Edge-based compartmental model for SIR disease on a static network.  Superficially, these models appear quite different, but we will see that a simple change of variables shows that they are actually very closely related.  This close relation will allow us to combine mixtures of the models.

\subsubsection{The mass action model}
\begin{figure}
\begin{center}
\raisebox{-0.5\height}{\begin{tikzpicture}

			\node[box, fill=colorS!60] (S1) at (-1,0.0) {$S$};
    			\node[box, fill=colorI!60] (I1) at (1,0.0) {$I$};
    			\node[box, fill=colorR!60] (R1) at (3,0) {$R$};
    			\path [->, above] (S1) edge node {$\beta SI$} (I1);
    			\path [decay, above ] (I1) -- node {$\gamma I$} (R1);			
                        \node [box, fill = colorI!60] (xi) at (2, -3) {$\xi$};
			\node (artificial) at (0,-3) {};
                        \path [->, above] (artificial) edge node {$\beta I$}   (xi);
		\end{tikzpicture}} \hfill
  \raisebox{-0.5\height}{\begin{tikzpicture}
    \node[box, fill=colorS!60] (S) at (-1,1) {$S$};
    \node[box, fill=colorI!60] (I) at (1,1) {$I$};
    \node[box, fill=colorR!60] (R) at (3,1) {$R$};
    \path [->] (S) edge node {} (I);
    \path [decay] (I) --(R) node [midway, above] {$\gamma I$} (R);
    \node[box, fill=colorS!30] (phiS) at ( -1,-2) {$\phi_S$};
    \node[box, fill=colorI!30] (phiI) at ( 1,-2) {$\phi_I$};
    \node[box, fill=colorR!30] (phiR) at ( 3,-2) {$\phi_R$};
    \begin{scope}[on background layer]
      \node[bigbox = {$\theta$}, fit=(phiS)(phiI)(phiR), fill=colorS!15] (Theta) {};
    \end{scope}
    \node[box,fill=colorI!60] (OmTheta) at (1,-4.5) {$1-\theta$};
    \path [->] (phiS) edge node {} (phiI);
    \path [decay] (phiI) -- (phiR) node [above, midway] {$\gamma\phi_I$};
    \path [->, right, near end] (phiI) edge node {$\tau\phi_I$} (OmTheta);
  \end{tikzpicture}}
\end{center}
\caption{(top left) A flow diagram demonstrating the standard model of mass action SIR dynamics.  (right) Two flow diagrams showing the EBCM model of network SIR dynamics.  (bottom left) A flow diagram, that when combined with the diagram above it allows for a system of equations similar to the EBCM model. }
\label{fig:basic_models}
\end{figure}
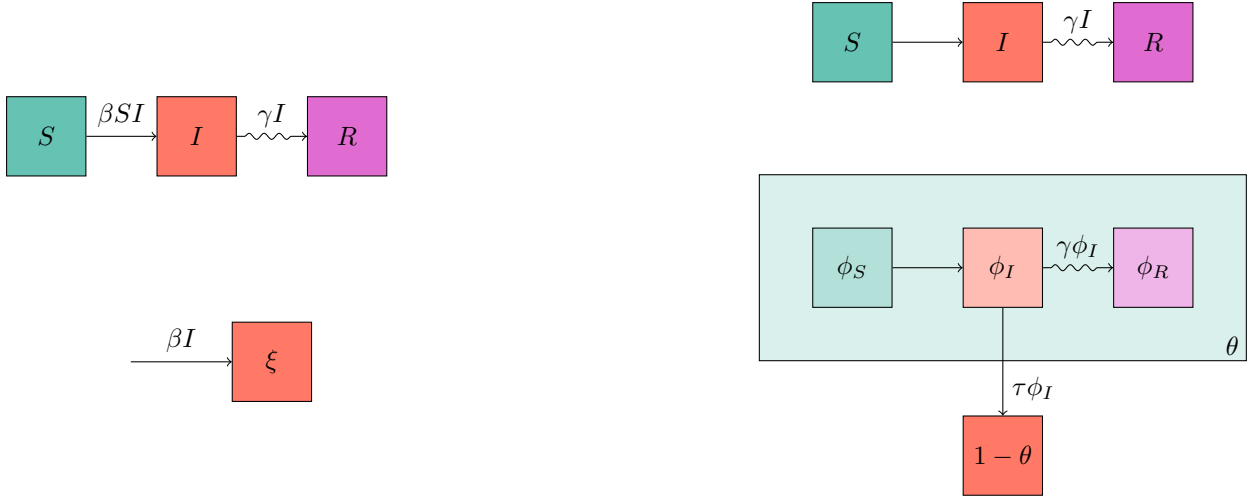

We begin with the mass action SIR model.  In this model, each individual causes new transmissions at rate $\beta$.  The recipient of the transmission is randomly chosen from the population.  If the recipient is susceptible, then she becomes infected and begins transmitting as a Poisson process with rate $\beta$.   She also recovers as a Poisson process with rate $\gamma$.  Once recovered, she remains immune to future infection.  The diagram in the top left of figure~\ref{fig:basic_models} leads us to the equations
\begin{subequations}
\label{eqn:MA}
\begin{align}
\dot{S} &= -\beta I S\\
\dot{I} &= \beta I S - \gamma I\\
\dot{R} &= \gamma I
\end{align}
\end{subequations}
with initial conditions
\begin{align*}
S(0) &= 1-\rho\\
I(0) &= \rho\\
R(0) &= 0 \, .
\end{align*}

A comparison of a stochastic simulation with solutions of system~\ref{eqn:MA} is given on the left of figure~\ref{fig:basic_model_solns}.  The simulations were performed with a population of 10000 individuals with $\beta=10$ and $\gamma=1$.  At $t=0$, \ $5\%$ of the population was randomly chosen to be infected ($\rho=0.05$).  The fit is excellent.

\subsubsection{Edge-based compartmental model}
For the simple network-based model we assume that a function $P(k)$ is known which gives the probability a randomly chosen individual has degree $k$, its number of partners.  If we assume that partnerships are randomly chosen, then the probability a random partner of a random individual has degree $k$ is $P_n(k) = k P(k)/\ave{K}$ where $\ave{K}$ is the average degree.  

This model is different from the mass action model because we assume individuals transmit along each partnership with rate $\tau$, and there is no global transmission.  As before, if the recipient is susceptible he becomes infected, recovering to an immune state with rate $\gamma$.    We follow the two diagrams on the right of figure~\ref{fig:basic_models} to get the Edge-based compartmental model (EBCM) for a system with an arbitrary initial fraction infected~\cite{miller:volz, miller:ebcm_overview, miller:initial_conditions}\cite[chapter~6]{EoNbook}
\begin{subequations}
\label{eqn:EBCM}
\begin{align}
\dot{\theta} &= - \tau \phi_I\\
\phi_I &= \theta - (1-\rho)\frac{\psi'(\theta)}{\ave{K}} - \frac{\gamma}{\tau} (1-\theta) \\
S&= (1-\rho)\psi(\theta)\\
I &= 1-S-R\\
\dot{R} &= \gamma I
\end{align}
\end{subequations}
with initial conditions
\begin{align*}
\theta(0)&=1\\
\end{align*}
where $\psi(x) = \sum_k P(k) x^k$ is the probability generating function of the degree distribution.  Here $\theta(t)$ represents the probability that at time $t$ a random partner of a randomly chosen individual $u$ has not transmitted to $u$.  These equations have been proven to be correct for random networks of given degree distribution in the infinite network size limit so long as the second moment $\ave{K^2}$ is finite~\cite{janson:SIRproof} (and under stronger assumptions by~\cite{decreusefond:volz_limit}).  

We briefly outline a derivation of these equations.  We divide $\theta$ into three parts: the probability a partner $v$ is susceptible and has not transmitted to $u$, the probability $v$ is infected and has not transmitted to $u$, and the probability $v$ is recovered and did not transmit to $u$.  We can solve for $\phi_I$ in terms of $\theta$.  We start with $\phi_I = \theta-\phi_S-\phi_R$.  It can be shown that $\phi_S = \phi_S(0) \psi'(\theta)/\ave{K}$ and $\phi_R = \gamma(1-\theta)/\tau + \phi_R(0)$.  Using our assumptions about the disease introduction at $t=0$, we know $\phi_S(0)=1-\rho$ and $\phi_R(0)=0$.  Thus $\phi_I = \theta - (1-\rho) \psi'(\theta)/\ave{K} - \gamma(1-\theta)/\tau$.  More details (including a subtle detail explaining why we can ignore transmissions from $u$ to $v$) is found in~\cite{miller:ebcm_overview} and~\cite[chapter 6]{EoNbook}.

\begin{figure}
\begin{center}
\includegraphics[width=0.4\textwidth]{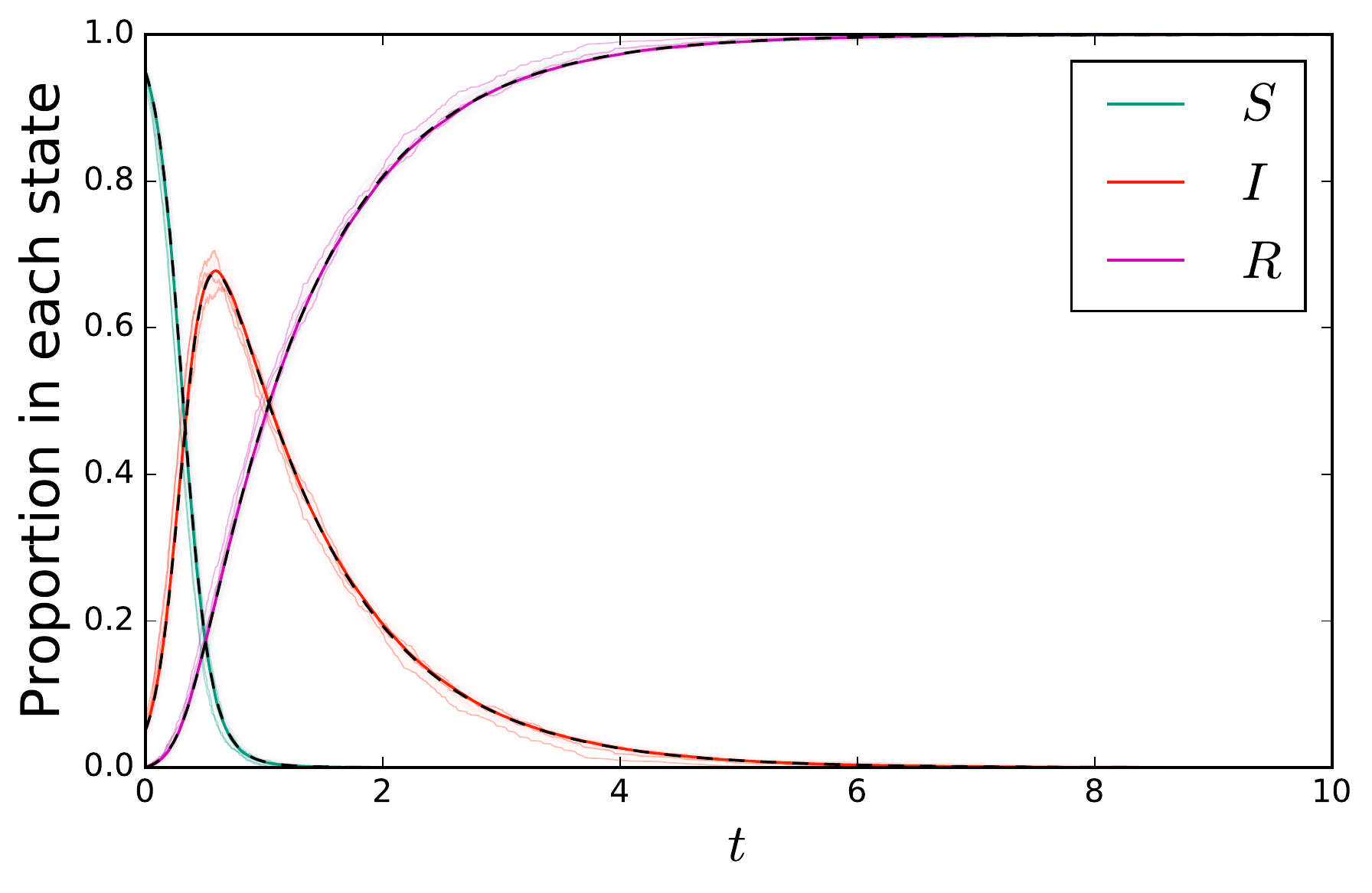} \hfill \includegraphics[width=0.4\textwidth]{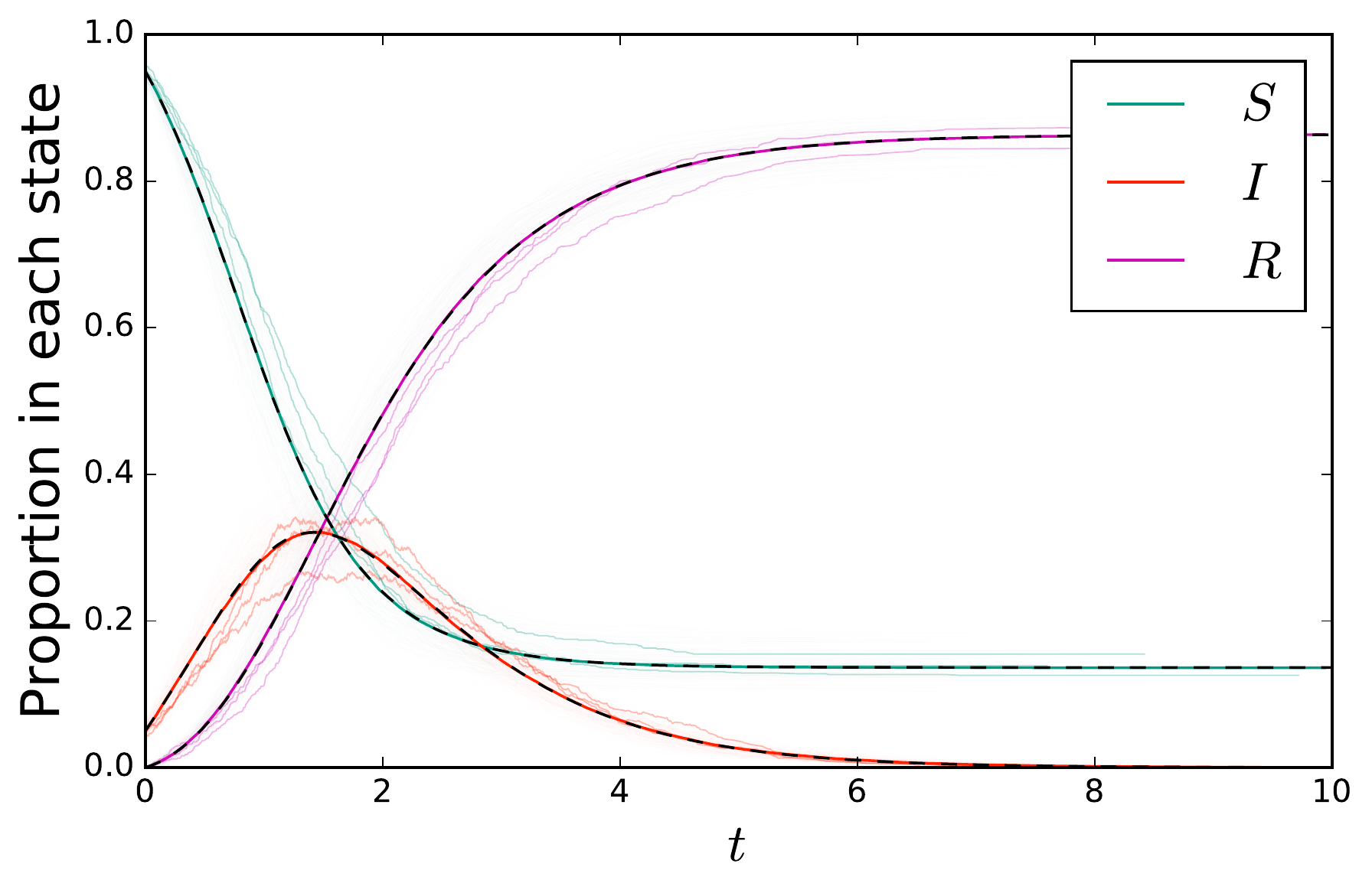}
\end{center}
\caption{(left) A comparison of stochastic agent-based simulation with ODE solution for the mass action model.  (right) A comparison of stochastic agent-based simulation with ODE solution for the network-based model.  In both cases, dashed curve denotes the solution to the ODE.  A cloud made up of $200$ simulations in populations of $1000$ individuals is shown lightly in color, with $3$ of these simulations highlighted.  An additional simulation in a population of $100000$ individuals is shown in darker color, almost exactly matching the ODE solution.}
\label{fig:basic_model_solns}
\end{figure}

A comparison of a stochastic simulation with solutions of system~\ref{eqn:EBCM} is given on the right of figure~\ref{fig:basic_model_solns}.  The network has $P(2)=P(4)=0.5$, so $\psi(x) = (x^2+x^4)/2$.  The simulations were performed with a population of 10000 individuals with $\tau=2$ and $\gamma=1$.  At $t=0$, a fraction $\rho$ was randomly chosen to be infected.

\subsubsection{Model similarity}
These two models~\eqref{eqn:MA} and~\eqref{eqn:EBCM} appear quite dissimilar in structure, which would make it difficult to derive a combined model.  However, we can modify the mass action model, making it easier to join the two into a common framework.

We define $\xi = \beta \int_0^t I(\hat{t}) \, \mathrm{d}\hat{t}$ to be the expected number of transmissions a random individual has received by time $t$ in the mass action model.  Using an integrating factor for the $\dot{S}$ equation, we see
\[
\dot{S} + \beta I S =0
\]
becomes
\[
\diff{}{t} \left(S e^{\beta \int_0^t I \, \mathrm{d}\hat{t}}\right) = 0
\]
From this it follows that 
\[
S(t)= S(0) e^{-\xi(t)} = (1-\rho)e^{-\xi(t)}
\]
and $\dot{R}=\gamma I$.  Finally substituting for $S(t)$ in $\dot{S} = -\beta I S$ gives us $\dot{\xi}=\beta\xi$.  As we assume $R(0)=0$, we have
\begin{subequations}
\label{eqn:xi}
\begin{align}
S &= (1-\rho)e^{-\xi}\\
I &= 1-S-R\\
\dot{R} &= \gamma I\\
\dot{\xi} &= \beta I = \beta\left[1-(1-\rho)e^{-\xi} - \frac{\gamma \xi}{\beta}\right]
\end{align}
\end{subequations}
with $\xi(0)=0$.  The variable $\xi$ plays a similar role to $\theta$ (more precisely $e^{-\xi}$ plays a similar role to $\psi(\theta)$).  Note that as for $\theta$ in system~\eqref{eqn:EBCM} we have an ODE for $\xi$ in terms of just $\xi$.  We could simplify this model further by noting that $\dot{R} = \gamma \dot{\xi}/\beta$ and so $R = \gamma \xi/\beta$.

In system~\eqref{eqn:xi}, we interpret $\xi$ as the expected number of transmissions an individual has received in the mass action population (the first of which causes infection).  An individual that expects to receive $\xi$ randomly distributed events has in fact received $0$ with probability $e^{-\xi}$.  Thus we could directly derive $S(t) = S(0) e^{-\xi(t)}$ without using system~\eqref{eqn:MA}.   We can interpret the equation for $\dot{\xi}$ as stating that the rate at which a random individual expects to receive a transmission is the transmission rate $\beta$ times the fraction of infected individuals.  This is demonstrated in the two diagrams on the left of figure~\ref{fig:basic_models}.  Among the advantages of this approach is that it is much easier to derive a final size relation from~\eqref{eqn:xi} than from~\eqref{eqn:MA} (see below).

\subsubsection{$\Ro$}
Often $\Ro$ is defined as the expected number of infections caused when a single infected individual is introduced into a fully susceptible population.  This definition is acceptable in a completely homogeneous, fully mixed population.  However, when the population exhibits heterogeneity, or partnerships have non-negligible duration, this definition breaks down --- if we want $\Ro$ to accurately capture aspects of the transmission in the early stages then it needs to accurately represent the typical individual infected early in the spread.  If some individuals are more likely to become infected, or if partnerships are long-lasting so that the infector of individual $v$ cannot be reinfected by $v$, then the typical introduced infection will cause a different number of infections than the typical individual infected early in the epidemic.  In defining $\Ro$ we must determine what the distribution of new infections settles down to early in the epidemic rather than what the distribution of introduced infections looks like~\cite{diekmann:R0,diekmann2009construction}.

In the mass action model, we assume that all infected individuals transmit at rate $\beta$, and the recipient is chosen uniformly at random from the population.  Early in the epidemic, each transmission reaches a susceptible individual, and successfully causes infection.  Infected individuals recover with rate $\gamma$, having an average duration of $1/\gamma$.  The expected number of infections caused by an individual infected early in the epidemic is thus $\Ro=\beta/\gamma$. This result is no different than what we would obtain by defining $\Ro$ in terms of a single introduced infection.

In the network case however, the $\Ro$ calculation is more delicate.  Early in the epidemic, a newly infected individual is infected with probability proportional to its degree, so the early infections have degree $k$ with probability $P_n(k) = kP(k)/\ave{K}$.  Further, these individuals will not be able to infect their infector.  Thus they have $k-1$ susceptible partners.  The probability of transmitting along a partnership prior to recovering is $\tau/(\tau+\gamma)$.  Thus the expected number of new infections caused is $\Ro = \sum_k k P(k) (k-1)\tau/(\tau+\gamma)\ave{K} = \ave{K^2-K}\tau/(\tau+\gamma)\ave{K}$.

\subsubsection{Final size relations}
For the mass action model, there is a well-known final size relation.  However, the derivation given is often quite circuitous (involving dividing $\dot{I}$ by $\dot{S}$ and integrating the result to find $I$ as a function of $S$.  We find that system~\eqref{eqn:xi} makes the derivation straightforward.  We have $R(t)  = \gamma \xi(t)/\beta$.  However, we also know that $R(\infty) = 1- S(\infty) = 1- S(0)e^{-\xi(\infty)}$.  Substituting for $\xi$ in the exponential gives 
\begin{equation}
R(\infty) = 1- S(0) e^{-\beta R(\infty)/\gamma},
\end{equation}
 which is the well-known final size relation~\cite{kermack}.  Alternately, we can write $\xi(\infty) = $.  The easiest way to solve this is through iteration, starting with a guess that $R(\infty)=0$ (equivalently $\xi(0)=0$), plugging it in and iteratively finding improved approximations of $R(\infty)$.

Similarly, the final size derivation of the EBCM equations is also straightforward.  We have $\phi_I(\infty) = 0$, so 
\begin{equation}
\theta(\infty) = \phi_S(0) \psi'(\theta(\infty))/\ave{K} - \gamma (1-\theta(\infty))/\tau
\label{eqn:EBCM_final}
\end{equation}
(taking $\phi_R(0)=0$).  We can also solve this implicit equation for $\theta(\infty)$ through iteration, starting with $\theta(0)=1$.  This relation is also widely known, though the derivation is typically done by a self-consistency argument in the final state~\cite{newman:spread}.  

In general, it is possible to generate a final size relation for an SIR model if the probability a random individual $v$ would transmit to $u$ is independent of the time at which $v$ is infected (here we do not care whether $u$ might have already received a transmission from another source)~\cite{kenah:networkbased,kenah:EPN}.  Thus rather than simulating a disease spreading through a network we can in principle generate a directed graph \emph{a priori} in which an edge from $v$ to $u$ means that if $v$ becomes infected then it will transmit to $u$.  Given this directed graph, and some introduced infections, an individual is infected iff there exists a directed path from an index individual to the individual of interest . More generally we can calculate the proportion infected in the limit of an infinitesimally small introduction as the proportion of the population with a giant ``in-component''~\cite{kenah:EPN,broder,dorogovtsev}.  With this in mind, the implicit equations derived can be interpreted as follows: Given a value of $\xi(\infty)$ or $\theta(\infty)$ we can infer the number and degrees of the individuals infected.  By knowing how many individuals become infected (and their degrees) we can infer how many transmissions occur.  That is, we can infer $\xi(\infty)$ or $\theta(\infty)$.  These must be consistent.  This yields the consistency relations above, where given $\xi$ or $\theta$ on the right hand side the inferred value appears on the left.

It is instructional to recognize that with appropriate initial condition the iteration corresponds directly to solving the dynamics of a particular discrete-time (Reed--Frost) disease model.\footnote{For example, with~\eqref{eqn:EBCM_final}, if we take $\theta_0=1$ and $\phi_S(0)=1-\rho$, then iteration directly gives us $\theta_1$, the value after one generation from which we can infer the number infected in the first generation.  Repeating, we can capture the entire dynamics.}  Each iteration gives the successive generation's size. Thus iterating until convergence gives the result as the number of generations tends to $\infty$.

If however the timing of $v$'s infection affects whether it would transmit to $u$ (as may happen if there are seasonal effects or if the number of other infections in the same timestep somehow influences $v$'s likelihood of transmitting), then we cannot derive a final size relation in the implicit way we have done it here.  Instead, we must solve the dynamical equations.  This represents the fact that we cannot predict the number of transmissions $v$ has caused until we know the time (and possibly the state) at which $v$ becomes infected.  

In later sections we will see that it is often easier to directly derive the consistency relation yielding the final size.

\subsection{The basic combined model}
Our first new model has a mass action model for global interactions combined
with transmission along a sexual network.

Guided by the models above, we develop the combined model.  We take $\theta$ to be the probability a random partner of an individual $u$ has not transmitted to $u$ (given that we ignore transmissions from $u$ to its partners), while we define $\xi(t)$ to be the expected number of transmissions $u$ has received by time $t$ through mass action transmission.

We assume that individuals have four possible statuses to capture our observations that sexual transmission may be much longer lasting.  They begin susceptible $S$.  Through either sexual or mass action interactions they become infectious.  The infectious phase is made up of two stages: the initial stage $I_1$ (in which they are infectious through global and sexual transmission) and the later stage $I_2$ (in which they are infectious only through sexual transmission).  They later move into a recovered stage $R$.  We assume that the transition rate from $I_1$ to $I_2$ is $\gamma_1$ and the rate from $I_2$ to $R$ is $\gamma_2$.  The symbols $S$, $I_1$, $I_2$, and $R$ will be used to denote both the stage and the fraction of individuals in that stage so $S+I_1+I_2+R=1$.  We finally assume that the mass action transmission rate is $\beta$, while the sexual transmission rate of the $I_1$ phase is $\tau_1$ and the $I_2$ phase is $\tau_2$.  To be clear about how these rates are normalized, $\beta$ is a rate of transmission to all other individuals, while $\tau_1$ and $\tau_2$ are rates of transmission to a specific partner.   

If we take $\tau_2=0$ (or $\gamma_2=\infty$), then this model becomes equivalent to a model of~\cite{ball:network_eqns}, although the model structure appears different (the equivalence can be showed through techniques in~\cite{miller:equivalence}).

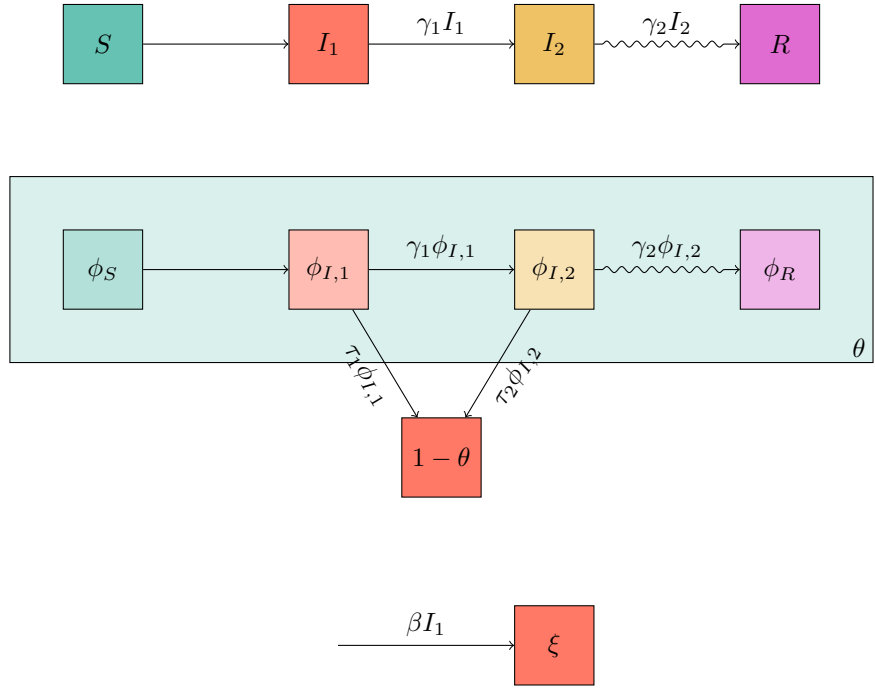
\begin{figure}
\begin{center}
\begin{tikzpicture}
    \node[box, fill=colorS!60] (S) at (0,1) {$S$};
    \node[box, fill=colorI!60] (I1) at (3,1) {$I_1$};
    \node[box, fill=colorI2!60] (I2) at (6,1) {$I_2$};
    \node[box, fill=colorR!60] (R) at (9,1) {$R$};
    \path [->] (S) edge node {} (I1);
    \path [->, above] (I1) edge node {$\gamma_1 I_1$} (I2);
    \path [decay] (I2) --(R) node [midway, above] {$\gamma_2 I_2$} (R);
    \node[box, fill=colorS!30] (phiS) at ( 0,-2) {$\phi_S$};
    \node[box, fill=colorI!30] (phiI1) at ( 3,-2) {$\phi_{I,1}$};
    \node[box, fill=colorI2!30] (phiI2) at ( 6,-2) {$\phi_{I,2}$};
    \node[box, fill=colorR!30] (phiR) at ( 9,-2) {$\phi_R$};
    \begin{scope}[on background layer]
      \node[bigbox = {$\theta$}, fit=(phiS)(phiR), fill=colorS!15] (Theta) {};
    \end{scope}
    \node[box,fill=colorI!60] (OmTheta) at (4.5,-4.5) {$1-\theta$};
    \path [->] (phiS) edge node {} (phiI1);
    \path [->, above] (phiI1) edge node {$\gamma_1 \phi_{I,1}$} (phiI2);
    \path [decay] (phiI2) -- (phiR) node [above, midway] {$\gamma_2\phi_{I,2}$};
    \path [->,below, sloped] (phiI1) edge node {$\tau_1\phi_{I,1}$} (OmTheta);
    \path [->,below, sloped] (phiI2) edge node {$\tau_2\phi_{I,2}$} (OmTheta);
    \node [box, fill = colorI!60] (xi) at (6, -7) {$\xi$};
    \node (artificial) at (3,-7) {};
    \path [->, above] (artificial) edge node {$\beta I_1$}   (xi);
  \end{tikzpicture}
\end{center}
\caption{Flow diagrams leading to system~\eqref{eqn:combined}.  The top diagram shows transitions of individuals between the states.  The middle diagram shows transitions of the status of a partner $v$ of a randomly chosen individual $u$ as well as whether $v$ has transmitted to $u$ (all ignoring transmissions to $u$ to $v$).  The bottom diagram shows the change in $\xi$ and how it relates to $I_1$.  These diagrams are related together by the dependence of the flux into $\xi$ on $I_1$ and on the fact that we can express $S$ and $\phi_S$ in terms of $\xi$ and $\theta$.}
\label{fig:combined}
\end{figure}

Figure~\ref{fig:combined} shows the corresponding flow diagrams for the combined model.

We first look for equations for $S$, $I_1$, $I_2$, and $R$.  We have $S(t) = (1-\rho) e^{-\xi(t)}\psi(\theta(t))$, that is $S$ is the probability a random individual was not initially infected and has not yet received a transmission.  Once this is set, we use the top diagram in figure~\ref{fig:combined} to find equations for the other variables.  We have $\dot{I}_2 = \gamma_1 I_1 - \gamma_2 I_2$ and $\dot{R} = \gamma_2 I_2$.  This leaves $I_1 = 1-S-I_2-R$.  

We now develop the equations for $\xi$ and $\theta$.  We begin with $\xi$, following figure~\ref{fig:combined}.  The equation is simply $\dot{\xi} = \beta I_1$, and the initial condition is $\xi(0)=0$.  

For $\theta$, we start with the observation that the probability a partner of $u$ is susceptible is $\phi_S = (1-\rho)e^{-\xi}\sum_k P_n(k) \theta^{k-1} = (1-\rho) e^{-\xi} \psi'(\theta)/\ave{K}$.  The remaining equations are found through figure~\ref{fig:combined}: We define $\phi_{I,1}$ to be the probability a partner is in the first infectious stage and has not transmitted to $u$.  We similarly define $\phi_{I,2}$ to be the probability the partner is in the second stage and has not transmitted to $u$.  We finally define $\phi_R$ to be the probability the partner is in the recovered stage and did not transmit to $u$.  Then $\theta = \phi_S + \phi_{I,1} + \phi_{I,2} + \phi_R$.  We arrive at $\dot{\phi}_{I,2} = \gamma_1 \phi_{I,1} - (\gamma_2+\tau_2)\phi_{I,2}$: the term with $\gamma_1$ denotes movement from the first to the second infectious stage (before transmitting to $u$).  The $\gamma_2$ term represents movement from the second infectious stage to the recovered stage (before transmitting to $u$).  The $\tau_2$ term represents the first transmission to $u$ occurring.  Similarly, we have $\dot{\phi}_R = \gamma_2 \phi_{I,2}$.  Our equation for $\theta$ is $\dot{\theta} = -\tau_1 \phi_{I,1} - \tau_2 \phi_{I,2}$.  We finally write $\phi_{I,1} = \theta- \phi_S - \phi_{I,2} - \phi_R$.  Note that in the previous model we could express $\phi_R$ as a simple multiple of $1-\theta$.  As the fluxes into the compartments are no longer proportional, we cannot do this now.

Putting this all together we have
\begin{subequations}
\label{eqn:combined}
\begin{align}
S &= (1-\rho) e^{-\xi} \psi(\theta)\\
I_1 &= 1- S - I_2 - R\\
\dot{I}_2 &= \gamma_1 I_1 - \gamma_2 I_2\\
\dot{R} &= \gamma_2 I_2\\
\dot{\xi} &= \beta I_1\\
\dot{\theta} &= -\tau_1 \phi_{I,1} - \tau_2 \phi_{I,2}\\
\phi_S &= (1-\rho)e^{-\xi}\frac{\psi'(\theta)}{\ave{K}}\\
\phi_{I,1} &= \theta - \phi_S - \phi_{I,2} - \phi_R\\
\dot{\phi}_{I,2} &= \gamma_1 \phi_{I,1} - (\gamma_2+\tau_2)\phi_{I,2}\\
\dot{\phi}_R &= \gamma_2 \phi_{I,2}
\end{align}
\end{subequations}

\begin{figure}
\begin{center}
\includegraphics[width=0.5\columnwidth]{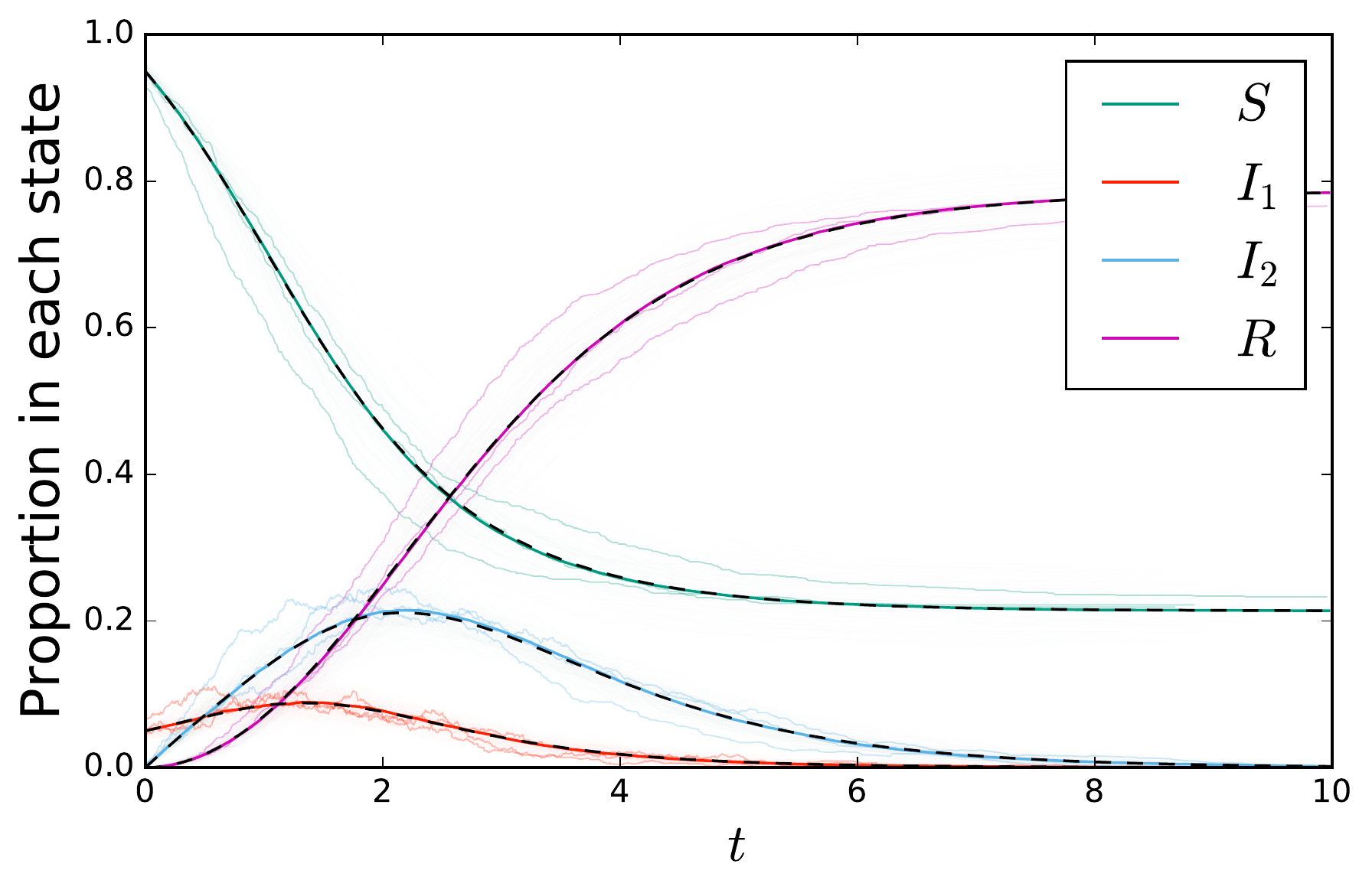}
\end{center}
\caption{A comparison of simulation with solutions to system~\eqref{eqn:combined} for network-based transmission combined with mass-action transmission.  The colored cloud consists of $200$ simulations in a population of $1000$ individuals, with three highlighted.  A darker solid colored curve shows a simulation in a population of $100000$.  This is almost exactly overlain by the solution to the ODE system~\eqref{eqn:combined} shown in a black dashed curve.}
\label{fig:combined_soln}
\end{figure}

Figure~\ref{fig:combined_soln} compares simulated epidemics in a population of 10000.  The transmissions occur in a mass action manner as in the mass action model above, but also in across a sexual network with $\psi(x) = (x^2+x^4)/2$.  The infectious period is divided into two stages.  The first infectious stage $I_1$ transmits through both mechanisms with $\tau_1=1$, \ $\beta=1$, and $\gamma_1 = 3$.  The second stage is longer lasting, but only infectious through the sexual network, with $\tau_2=0.5$ and $\gamma_2=1$.  

\subsubsection{$\Ro$}

We will use two different derivations of $\Ro$ for this population.  This is because each is more appropriate for different generalizations presented below.   In both cases, we distinguish individuals by their ``generation'', that is, the number of sequential transmission events from an initial infection needed to infect an individual.  

In our first derivation, we distinguish those individuals infected through a mass action transmission from those infected through a sexual transmission.  We use $N_{ma}(g)$ and $N_s(g)$ to denote the expected number of each in generation $g$.  Early in the spread, we can ignore depletion of susceptible individuals.  We first consider the number of transmissions an individual causes through the mass action route.  Regardless of how the individual is infected, the transmission rate is $\beta$ and the rate of departing the infectious phase is $\gamma_1$.  Thus the expected number of mass-action transmissions is $R_{ma}=\beta/\gamma_1$.  

The average individual infected through a mass action transmission has $\ave{K}$ susceptible partners, while the average individual infected through a sexual transmission has $\ave{K^2-K}/\ave{K}$ susceptible partners.  The probability of no sexual transmission during the first infectious stage is $\gamma_1/(\tau_1+\gamma_1)$.  The probability of no sexual transmission during the second stage (given that no transmission occurred during the first) is $\gamma_2/(\tau_2+\gamma_2)$.  Thus the total probability of a sexual transmission is $1-(\gamma_1\gamma_2)/(\tau_1+\gamma_1)(\tau_2+\gamma_2)$.  Thus the expected number of sexual transmissions caused given that an individual is infected through a mass action transmission is $R_{s|ma} = \ave{K} (1- \gamma_1\gamma_2/(\tau_1+\gamma_1)(\tau_2+\gamma_2))$ and the expected number of sexual transmissions caused given that an individual is infected through a sexual transmission is $R_{s|s} = \ave{K^2-K}(1-\gamma_1\gamma_2/(\tau_1+\gamma_1)(\tau_2+\gamma_2))/\ave{K}$. 

Combining this, we see that
\[
\begin{pmatrix}
N_{ma}(g+1)\\
N_s(g+1)
\end{pmatrix}
= 
\begin{pmatrix}
R_{ma} & R_{ma}\\
R_{s|ma} & R_{s|s}
\end{pmatrix}
\begin{pmatrix}
N_{ma}(g)\\
N_s(g)
\end{pmatrix}
\]
At leading order, as $g \to \infty$, this will converge to
$c\lambda^g \vec{x}$ where $\lambda$ is the dominant eigenvalue
and $\vec{x}$ its eigenvector.  This dominant eigenvalue is $\Ro$.
\begin{align*}
\Ro &=\frac{ R_{ma} + R_{s|s} + \sqrt{(R_{ma}+R_{s|s})^2-4(R_{ma}R_{s|s}-R_{s|ma}R_{ma})}}{2}\\
&=\frac{ R_{ma} + R_{s|s} + \sqrt{(R_{ma}-R_{s|s})^2+4R_{s|ma}R_{ma}}}{2}\\
& = \frac{ R_{ma} + R_{s|s} + |R_{ma}-R_{s|s}|}{2} + \frac{R_{s|ma}R_{ma}}{|R_{ma}-R_{s|s}|} + \order\left(\frac{(R_{s|ma}R_{ma})^2}{|R_{ma}-R_{s|s}|^3}\right)
\end{align*}
If $R_{ma}=0$, this reduces to $R_{s|s}$ (and if there is no sexual transmission this reduces to $R_{ma}$).  If $R_{ma}R_{s|ma} \neq 0$, then this is guaranteed to be larger than the largest of $R_{ma}$ and $R_{s|s}$.  If $R_{ma}$ and $R_{s|s}$ are very close to one another, the expansion above is not well behaved.  Then the more appropriate expansion yields $\Ro \approx R_{ma} + \sqrt{R_{ma} R_{s|ma}}$.

For our second derivation, we make the observation that any variable which we expect to be proportional to the number of infections will behave like $\lambda^g$ for large $g$.  Rather than distinguishing by how an individual is infected, we simply count the number of infected individuals and the number of $SI$ partnerships forming in each generation.  We take $N(g)$ to denote the expected number infected in generation $g$, and $y(g)$ to denote the number of $SI$ partnerships in generation $g$.  We anticipate that these will grow at rate $\Ro$, and that if we know their values for one $g$, we can calculate their values for $g+1$.

To determine how many individuals are infected in generation $g+1$, we note that on average each individual in generation $g$ causes $R_{ma}$ infections through the mass action route, so there are $N(g)R_{ma}$ newly infected individuals through this route.  Through the sexual transmission route, we simply note that each edge transmits with probability $1-\gamma_1\gamma_2/(\tau_1+\gamma_1)(\tau_2+\gamma_2)= R_{s|m}/\ave{K}$.  So there are $y(g) R_{s|m}/\ave{K}$ new infections through sexual transmissions.  We can count $N(g)R_{ma} \ave{K}$ new SI partnerships resulting from transmission through the mass action route, and $y(g) (1-\gamma_1\gamma_2/(\tau_1+\gamma_1)(\tau_2+\gamma_2))(\ave{K^2-K}/\ave{K}) = y(g) R_{s|s}$ new SI partnerships through the sexual transmission route.  Thus
\[
\begin{pmatrix}
N(g+1)\\
y(g+1)
\end{pmatrix}
= 
\begin{pmatrix}
R_{ma} & R_{ma}\ave{K}\\
R_{s|m}/\ave{K} & R_{s|s}
\end{pmatrix}
\begin{pmatrix}
N(g)\\
y(g)
\end{pmatrix}
\]
It is straightforward to see that this has the same eigenvalues as above.  The dominant eigenvalue is still $\Ro$.

\subsubsection{Final Size Relation}
We can derive a final size relation.  We take $t \to \infty$ and note that all of the variables corresponding to active infectious states must go to zero.  Then
\begin{align*}
S(\infty) &= (1-\rho) e^{-\xi(\infty)}\psi(\theta(\infty))\\
\phi_S(\infty) &= (1-\rho) e^{-\xi(\infty)}\frac{\psi'(\theta(\infty))}{\ave{K}}
\end{align*}
We will express $\xi(\infty)$ and $\theta(\infty)$ in terms of $S(\infty)$ and $\phi_S(\infty)$.  This will lead to an implicit relation which can be solved.

It is straightforward to see that because $I_2(0)=0$ and $I_2(\infty)=0$ we have $\gamma_1 \int_0^\infty I_1 \, \mathrm{d}t = \gamma_2 \int_0^\infty I_2 \mathrm{d}t =$.  However $R(\infty) = \gamma_2 \int_0^\infty I_2 \mathrm{d}t$ and $\xi(\infty) = \beta \int_0^\infty  I_1 \mathrm{d}t$.  So we conclude that 
\[
\xi(\infty) = \frac{\beta}{\gamma_1} R(\infty) = \frac{\beta}{\gamma_1} [1-S(\infty)] = \frac{\beta}{\gamma_1} \left[ 1- (1-\rho) e^{-\xi(\infty)}\psi(\theta(\infty))\right]
\]

We now derive $\theta(\infty)$.  We first make the observation that $\theta(\infty) = \phi_S(\infty) + \phi_R(\infty)$.  Since $\phi_S(\infty) = (1-\rho) e^{-\xi(\infty)}\psi'(\theta(\infty))/\ave{K}$, we conclude $\phi_R(\infty) = \theta(\infty) - (1-\rho) e^{-\xi(\infty)}\psi'(\theta(\infty))/\ave{K}$.  

We now take the equation for $\dot{\theta}$ and integrate it, yielding $\theta(\infty) = 1 - \tau_1 \int_0^\infty \phi_{I,1} \mathrm{d}t - \tau_2 \int_0^\infty \phi_{I,2} \mathrm{d}t$.  We attack the integrals sequentially.  We  have $\dot{\phi}_R + \dot{\phi}_{I,2} = \gamma_1 \phi_{I,1} - \tau_2 \phi_{I,2}$.  Integrating this from $0$ to $\infty$ and noting that $\phi_{I,2}(0)=\phi_{I,2}(\infty)=0$ we conclude that 
\[
\int_0^\infty \phi_{I,1} \mathrm{d}t  = \frac{\phi_R(\infty)}{\gamma_1} + \frac{\tau_2 \int_0^\infty \phi_{I,2} \mathrm{d}t}{\gamma_1} 
\]
So we have 
\[
\theta(\infty) = 1 - \frac{\tau_1}{\gamma_1} \phi_R(\infty) - \left(\frac{\tau_1\tau_2}{\gamma_1} + \tau_2\right) \int_0^\infty \phi_{I,2} \mathrm{d}t
\]
Similarly since $\dot{\phi}_R = \gamma_2 \phi_{I,2}$ and $\phi_R(0)=0$, we have $\int_0^\infty \phi_{I,2} \mathrm{d} t= \frac{\phi_R(\infty)}{\gamma_2}$.  Thus
\begin{align*}
\theta(\infty) &= 1 - \left[\frac{\tau_1}{\gamma_1} + \frac{\tau_1\tau_2}{\gamma_1\gamma_2} + \frac{\tau_2}{\gamma_2}\right] \phi_R(\infty)  \\
  & = 1- \left[\frac{\gamma_1+\tau_1}{\gamma_1}\frac{\gamma_2+\tau_2}{\gamma_2}-1\right]\phi_R\\
 &=  1- \left[\frac{\gamma_1+\tau_1}{\gamma_1}\frac{\gamma_2+\tau_2}{\gamma_2}-1\right]\left[ \theta(\infty) - \frac{(1-\rho) e^{-\xi(\infty)}\psi'(\theta(\infty))}{\ave{K}} \right]
\end{align*}

So we finally have
\begin{align*}
\xi(\infty) &= \frac{\beta}{\gamma_1} \left[ 1- (1-\rho) e^{-\xi(\infty)}\psi(\theta(\infty))\right]\\
\theta(\infty) &= 1- \left[\frac{\gamma_1+\tau_1}{\gamma_1}\frac{\gamma_2+\tau_2}{\gamma_2}-1\right]\left[ \theta(\infty) - \frac{(1-\rho) e^{-\xi(\infty)}\psi'(\theta(\infty))}{\ave{K}} \right]
\end{align*}
We can solve this system iteratively and substitute the result into our expression for $S(\infty)$ to find the final size of the epidemic.

\paragraph{Direct derivation}
The equation for $\theta(\infty)$ can be rearranged to give 
\[
\theta(\infty) = \frac{\gamma_1}{\gamma_1+\tau_1}\frac{\gamma_2}{\gamma_2+\tau_2} + \left[ 1 - \frac{\gamma_1}{\gamma_1+\tau_1}\frac{\gamma_2}{\gamma_2+\tau_2} \right]\frac{(1-\rho)e^{-\xi}\psi'(\theta(\infty))}{\ave{K}}
\]
It turns out we can derive this relation directly.  

We first observe that $\gamma_i/(\tau_i+\gamma_i)$ is the probability an individual who enters the $i$th infectious stage moves to the next stage before transmitting.  Thus if we set $T = 1- \frac{\gamma_1}{\gamma_1+\tau_1}\frac{\gamma_2}{\gamma_2+\tau_2} $, then we can interpret $T$ as the probability that at least one of the two stages transmits.  Thus since $\theta(\infty)$ must be the probability that a sexual partner either would not transmit even if infected, or would transmit but does not become infected, we have
\[
\theta(\infty) = 1-T + T \frac{(1-\rho)e^{-\xi}\psi'(\theta(\infty))}{\ave{K}}
\]
Similarly we can directly derive $\xi(\infty)$.  This is the expected number of mass action transmissions an individual receives.  The expected number of mass action transmissions caused by each infected individual is $\beta/\gamma_1$.  So $\xi(\infty) = \beta[1- S(\infty)]/\gamma_1$.  Substituting $S=(1-\rho)e^{-\xi}\psi(\theta)$ into this we have
\[
\xi(\infty) = \frac{\beta}{\gamma_1} \left[ 1 - (1-\rho)e^{-\xi(\infty)} \psi(\theta(\infty))\right]
\]
and thus we have the final size relation.

\subsection{A vector-borne transmission}
We now consider transmission involving a vector component, which without loss of generality we will refer to as a mosquito.  We assume that the lifetime of mosquitos is short compared to the duration of the epidemic.  We will ignore seasonal effects on the mosquito lifespan, but it would be straightforward to incorporate.  We assume that there is no latent phase: when an individual or a mosquito becomes infected, it is immediately infectious.  We still assume two infectious phases in humans, with human to mosquito transmission possible only in the first phase.

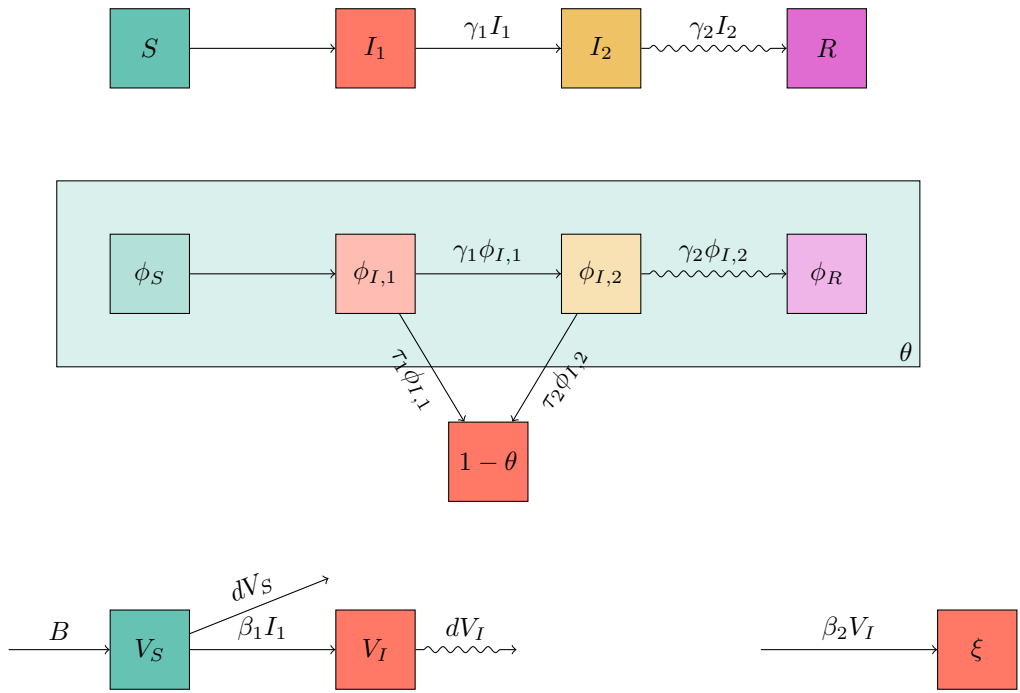
\begin{figure}
\begin{center}
\begin{tikzpicture}
    \node[box, fill=colorS!60] (S) at (0,1) {$S$};
    \node[box, fill=colorI!60] (I1) at (3,1) {$I_1$};
    \node[box, fill=colorI2!60] (I2) at (6,1) {$I_2$};
    \node[box, fill=colorR!60] (R) at (9,1) {$R$};
    \path [->] (S) edge node {} (I1);
    \path [->, above] (I1) edge node {$\gamma_1 I_1$} (I2);
    \path [decay] (I2) --(R) node [midway, above] {$\gamma_2 I_2$};
    \node[box, fill=colorS!30] (phiS) at ( 0,-2) {$\phi_S$};
    \node[box, fill=colorI!30] (phiI1) at ( 3,-2) {$\phi_{I,1}$};
    \node[box, fill=colorI2!30] (phiI2) at ( 6,-2) {$\phi_{I,2}$};
    \node[box, fill=colorR!30] (phiR) at ( 9,-2) {$\phi_R$};
    \begin{scope}[on background layer]
      \node[bigbox = {$\theta$}, fit=(phiS)(phiR), fill=colorS!15] (Theta) {};
    \end{scope}
    \node[box,fill=colorI!60] (OmTheta) at (4.5,-4.5) {$1-\theta$};
    \path [->] (phiS) edge node {} (phiI1);
    \path [->, above] (phiI1) edge node {$\gamma_1 \phi_{I,1}$} (phiI2);
    \path [decay] (phiI2) -- (phiR) node [above, midway] {$\gamma_2\phi_{I,2}$};
    \path [->,below, sloped] (phiI1) edge node {$\tau_1\phi_{I,1}$} (OmTheta);
    \path [->,below, sloped] (phiI2) edge node {$\tau_2\phi_{I,2}$} (OmTheta);
    \node [box, fill = colorI!60] (xi) at (11, -7) {$\xi$};
    \node (artificial) at (8,-7) {};
    \path [->, above] (artificial) edge node {$\beta_2 V_I$}   (xi);
    \node [box, fill = colorS!60] (VS) at (0, -7) {$V_S$} ;
    \node [box, fill = colorI!60] (VI) at (3, -7) {$V_I$} ;
    \node (artificial2) at (5,-7) {};
    \node (artificial3) at (-2,-7) {};
    \node (artificial4) at (2.5,-6) {};
    \path [->, above] (artificial3) edge node {$B$} (VS);
    \path [decay] (VI) --(artificial2) node [midway, above] {$d V_I$};
    \path [->, above] (VS) edge node {$\beta_1 I_1$} (VI);
    \path [->, above, sloped] (VS) edge node {$d V_S$} (artificial4);
  \end{tikzpicture}
\end{center}
\caption{Flow diagrams that lead to the governing equations for the combined sexual network and vector transmission routes.  These are very similar to figure~\ref{fig:combined}, except for the introduction of the vector component at the bottom.}
\label{fig:vector}
\end{figure}

The equations are very similar, except that we must handle the mosquitos as a separate compartment rather than assuming transmission is direct from human to human.  Mosquitos have a typical life span of $1/d$, and we assume a constant death rate of $d$ for adult mosquitos.  For the sexual transmission, we make the same assumptions as before.  We assume a constant influx (births) of new mosquitos $B$, measured in terms of mosquitos per unit time per person.

We introduce $V_S$ and $V_I$ to represent the number of susceptible and infected mosquitos per person.  The human to mosquito transmission rate is $\beta_1$, and the mosquito to human transmission rate is $\beta_2$.  We can write $\dot{V}_S$ to be the birth rate of new mosquitos minus the infection rate of susceptible mosquitos minus the death rate of susceptible mosquitos 
\[
 \dot{V}_S = B - dV_S - \beta_1 I_1 V_S 
\]
Similarly we have $\dot{V}_I = \beta_1 I_1 V_S - d V_I$.  When the births balance the deaths, we will have $B-dV_S = dV_I$.  

We set $\xi$ to be the expected number of transmissions received by a human from mosquitos.  Taking the other variables to be as before and following the flow diagram in figure~\ref{fig:vector}, we have 
\begin{subequations} \label{eqn:mos}
\begin{align}
S &= (1-\rho) e^{-\xi} \psi(\theta)\\
I_1 &= 1- S - I_2 - R\\
\dot{I}_2 &= \gamma_1 I_1 - \gamma_2 I_2\\
\dot{R} &= \gamma_2 I_2\\
\dot{V}_I &=\beta_1 I_1  V_S  - d V_I\\
\dot{V}_S &= B - d V_S - \beta_1 I_1 V_S\\
\dot{\xi} &= \beta_2 V_I\\
\dot{\theta} &= -\tau_1 \phi_{I,1} - \tau_2 \phi_{I,2}\\
\phi_S &=  (1-\rho)e^{-\xi}\frac{\psi'(\theta)}{\ave{K}} \\
\phi_{I,1} &= \theta -\phi_S - \phi_{I,2} - \phi_R\\
\dot{\phi}_{I,2} &= \gamma_1 \phi_{I,1} - (\gamma_2+\tau_2)\phi_{I,2}\\
\dot{\phi}_R &= \gamma_2 \phi_{I,2}
\end{align}
\end{subequations}

\begin{figure}
\begin{center}
\includegraphics[width=0.5\columnwidth]{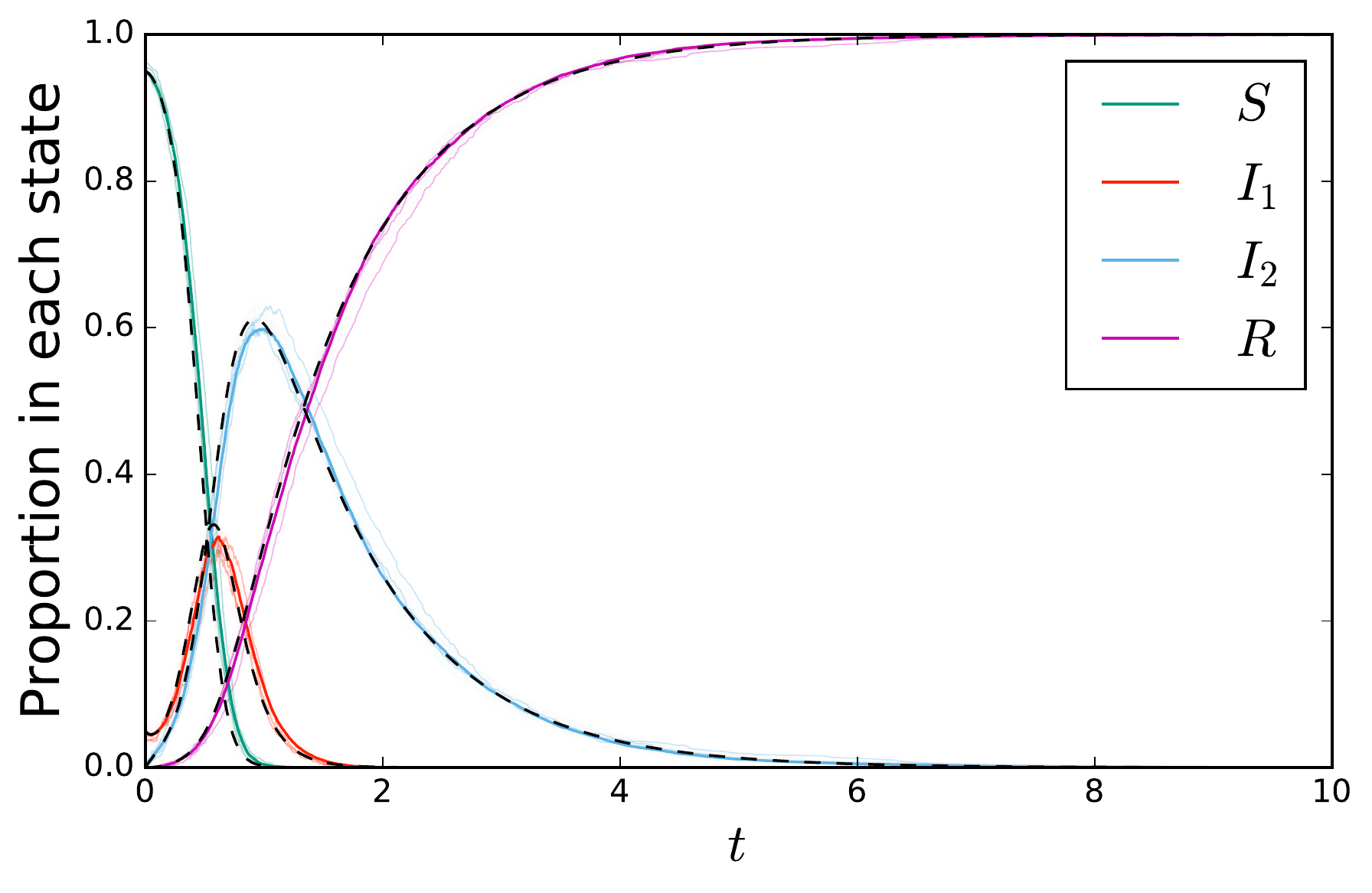}
\end{center}
\caption{A comparison of simulation with solutions to system~\eqref{eqn:mos}.  Because resolving individual mosquitos significantly increases calculation time, we run fewer calculations with smaller populations.  The cloud of simulations consist of $100$ simulations in a population of $1000$ individuals with typically $20$ mosquitos per individual.  Again $3$ are highlighted.  An additional solid colored curve is given for a population of $10000$ individuals.  The dashed black curve represents the solution to the ODEs of system~\eqref{eqn:mos}.}
\label{fig:mosquito_soln}
\end{figure}

Figure~\ref{fig:mosquito_soln} compares simulated epidemics in a population of 10000, with an additional mosquito population.   Each mosquito dies with rate $1$, and they are born at a rate such that on average there are 20 mosquitos per individual.  The transmissions can occur either through mosquitos or a sexual network with $\psi(x) = (x^2+x^4)/2$.  As before, the infectious period is divided into two stages.  The first infectious stage $I_1$ transmits through both mechanisms with $\tau_1=1$, \ $\beta_1= \beta_2=2$, and $\gamma_1 = 5$.  The second stage is longer lasting, but only infectious through the sexual network, with $\tau_2=0.3$ and $\gamma_2=1$.  

\subsubsection{$\Ro$}
Because of the similarity to the mass action model, we see that $R_{s|s}$ follows the same expression as before.  We can take $R_{s|mo}$ to correspond to $R_{s|ma}$ except that the infection is from a mosquito.  Its expression remains the same.  We assume that the mosquito population size is at equilibrium, $B/d$.  We can replace $R_{ma}$ with $R_{mo} = (B/d)(\beta_1/\gamma_1)(\beta_2/\gamma_2)$.  Then $\Ro$ is the leading eigenvalue of
\[
\begin{pmatrix}
R_{mo} & R_{mo}\\
R_{s|mo} & R_{s|s}
\end{pmatrix}
\]
and so it takes the same form as before, with the transmission through the mosquitos simply representing a delay in the infections rather than a fundamental difference in the number of transmissions an individual causes (early in the epidemic).

\subsubsection{Final size}
Given the similarity in the models~\eqref{eqn:combined} and~\eqref{eqn:mos} as well as the similarity in $\Ro$, we might expect a similar final size relation to emerge as in the previous model.  However, we cannot derive a final size relation in this case.  This is because there is no simple relation that allows us to calculate the number of infections caused through mosquitos based on knowing the total number of infections that occur.  If the epidemic has a sharp peak, some human to mosquito transmissions will go to an already infected mosquito, while if the epidemic is very broad and low very few infected mosquitos will bite an actively infected human.  So the total number of infected mosquitos depends on the dynamics of the epidemic, not just the total number of human infections.  We cannot uniquely determine how many mosquitos are infected based on the number of infected humans, thus we cannot uniquely determine how many humans become infected from those mosquitos.

If we take $\int_0^\infty V_I \mathrm{d} t$ as a given, we can derive $\xi$ and thus a final size relation, but this is a very strong assumption.  Alternately, we could assume that all human to mosquito transmissions go to a susceptible mosquito.  Then $\int_0^\infty V_I \mathrm{d}t = \beta_1 \int_0^\infty I(t) \mathrm{d}t / d \gamma_1$.  This would give an upper bound to the number of transmissions occurring, which in turn gives an upper bound to the epidemic size.  Note that if the epidemic is small or very slow, we would expect this to become a reasonable approximation, and so this will be a reasonable prediction of the epidemic size.

\subsection{Two overlapping networks}
\cite{miller:ebcm_overview}
We now consider a population in which the potentially transmitting contacts can be structured into two networks.  The individuals in the networks are the same, but the partnerships have different meanings.  The first network is the network of sexual partnerships, and the second network is a network of social interactions which can cause transmission.  As before we assume two infectious stages.  The first transmits with rate $\tau_1$ to sexual partners and $\beta$ to social partners.  The second transmits with rate $\tau_2$ to sexual partners and does not transmit to social partners.

\begin{figure}
\begin{center}
\begin{tikzpicture}
    \node[box, fill=colorS!60] (S) at (0,1) {$S$};
    \node[box, fill=colorI!60] (I1) at (3,1) {$I_1$};
    \node[box, fill=colorI2!60] (I2) at (6,1) {$I_2$};
    \node[box, fill=colorR!60] (R) at (9,1) {$R$};
    \path [->] (S) edge node {} (I1);
    \path [->, above] (I1) edge node {$\gamma_1 I_1$} (I2);
    \path [decay] (I2) --(R) node [midway, above] {$\gamma_2 I_2$} (R);

    \node[box, fill=colorS!30] (phiSse) at ( 0,-2) {$\phi^{se}_S$};
    \node[box, fill=colorI!30] (phiI1se) at ( 3,-2) {$\phi^{se}_{I,1}$};
    \node[box, fill=colorI2!30] (phiI2se) at ( 6,-2) {$\phi^{se}_{I,2}$};
    \node[box, fill=colorR!30] (phiRse) at ( 9,-2) {$\phi^{se}_R$};
    \begin{scope}[on background layer]
      \node[bigbox = {$\theta_{se}$}, fit=(phiSse)(phiRse), fill=colorS!15] (Thetase) {};
    \end{scope}
    \node[box,fill=colorI!60] (OmThetase) at (4.5,-4.5) {$1-\theta_{se}$};
    \path [->] (phiSse) edge node {} (phiI1se);
    \path [->, above] (phiI1se) edge node {$\gamma_1 \phi^{se}_{I,1}$} (phiI2se);
    \path [decay] (phiI2se) -- (phiRse) node [above, midway] {$\gamma_2\phi^{se}_{I,2}$};
    \path [->,below, sloped] (phiI1se) edge node {$\tau_1\phi^{se}_{I,1}$} (OmThetase);
    \path [->,below, sloped] (phiI2se) edge node {$\tau_2\phi^{se}_{I,2}$} (OmThetase);

    \node[box, fill=colorS!30] (phiSso) at ( 1.5,-7.5) {$\phi^{so}_S$};
    \node[box, fill=colorI!30] (phiIso) at ( 4.5,-7.5) {$\phi^{so}_I$};
    \node[box, fill=colorR!30] (phiRso) at ( 7.5,-7.5) {$\phi^{so}_R$};
    \begin{scope}[on background layer]
      \node[bigbox = {$\theta_{so}$}, fit=(phiSso)(phiRso), fill=colorS!15] (Thetaso) {};
    \end{scope}
    \node[box,fill=colorI!60] (OmThetaso) at (4.5,-10) {$1-\theta_{so}$};
    \path [->] (phiSso) edge node {} (phiIso);
    \path [decay] (phiIso) -- (phiRso) node [above, midway] {$\gamma_1\phi^{so}_I$};
    \path [->,right, near end] (phiIso) edge node {$\beta\phi^{so}_I$} (OmThetaso);
  \end{tikzpicture}
\end{center}
\caption{Flow diagrams that lead to the equations for a population in which interactions all lie within a sexual or a social network.  We assume that the infectious stage can be subdivided into two stages, the first of which transmits through either contact, and the second transmits only through sexual contact.}
\label{fig:two_networks}
\end{figure}
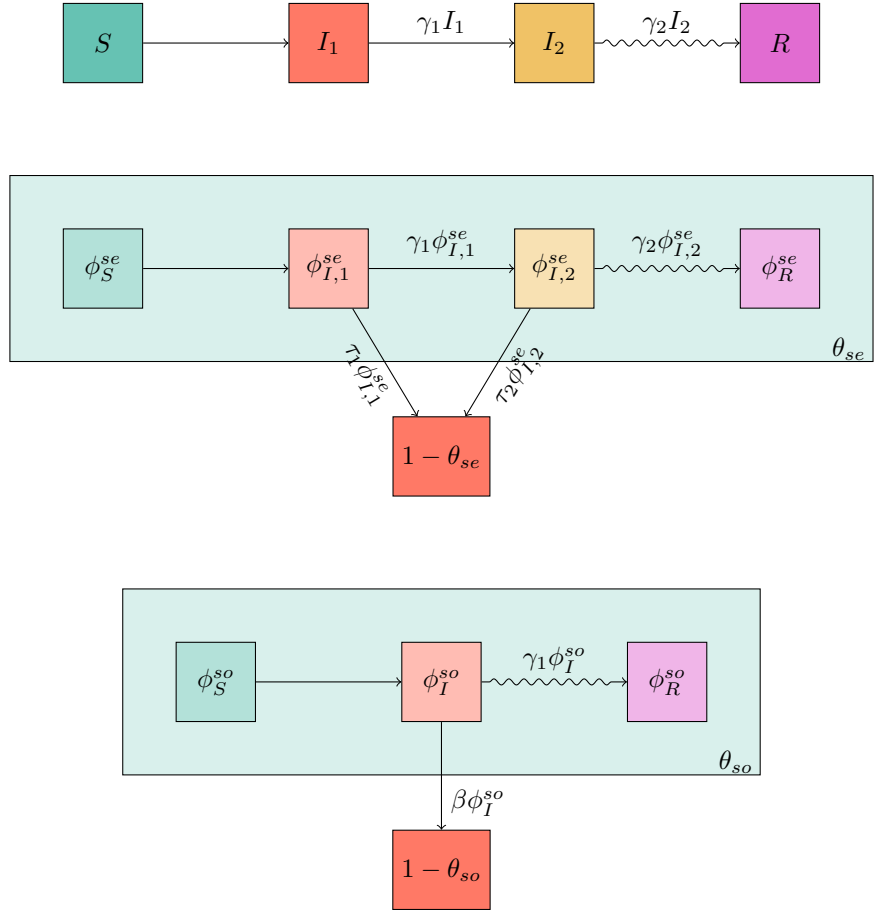

We assume a joint degree distribution $P(k_{se},k_{so})$ (that is, the number of sexual partners may correlate with the number of social partners) and define $\psi(x,y) = \sum_{k_{se},k_{so}} P(k_{se},k_{so}) x^{k_{se}}y^{k_{so}}$.  We define $\theta_{se}$ to be the probability that a sexual partnership has not yet transmitted to an individual $u$ (who is artificially prevented from transmitting) and $\theta_{so}$ to be the probability that a social partnership has not yet transmitted to $u$.  We use $\phi^{se}$ and $\phi^{so}$ depending on the type of partnership, with the subscript denoting the status of the partner.  As the second infectious stage is not infectious through social contacts, we combine the $\phi^{so}_{I,2}$ and $\phi^{so}_R$ into a single term $\phi^{so}_R$.  The diagrams in figure~\ref{fig:two_networks} give
\begin{subequations}
\label{eqn:two_networks}
\begin{align}
S &= (1-\rho) \psi(\theta_{se},\theta_{so})\\
I_1 &= 1- S - I_2 - R\\
\dot{I}_2 &= \gamma_1 I_1 - \gamma_2 I_2\\
\dot{R} &= \gamma_2 I_2\\
\dot{\theta}_{se} &= -\tau_1 \phi^{se}_{I,1} - \tau_2 \phi^{se}_{I,2}\\
\phi^{se}_{I,1} &= \theta - (1-\rho)\frac{\psi_x(\theta_{se},\theta_{so})}{\ave{K_{se}}} - \phi^{se}_{I,2} - \phi^{se}_R\\
\dot{\phi}^{se}_{I,2} &= \gamma_1 \phi^{se}_{I,1} - (\gamma_2+\tau_2)\phi^{se}_{I,2}\\
\dot{\phi}^{se}_R &= \gamma_2 \phi^{se}_{I,2}\\
\dot{\theta}_{so} &= -\beta \phi^{so}_{I,1}\\
\phi^{so}_{I,1} &= \theta - (1-\rho)\frac{\psi_y(\theta_{se},\theta_{so})}{\ave{K_{so}}}  - \phi^{so}_R\\
\dot{\phi}^{so}_R &= \gamma_1 \phi^{so}_{I,1}
\end{align}
\end{subequations}

\begin{figure}
\begin{center}
\includegraphics[width=0.5\columnwidth]{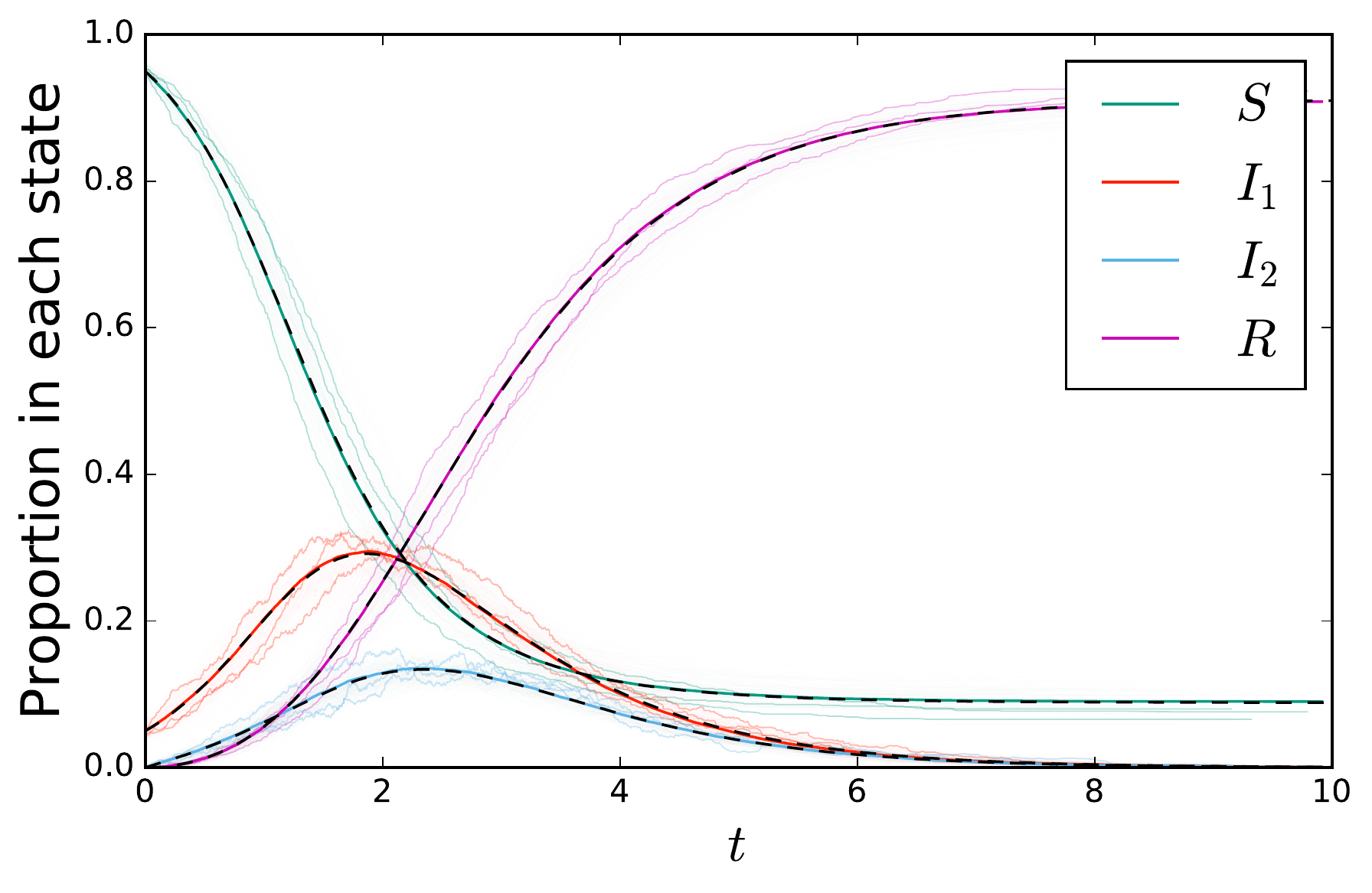}
\end{center}
\caption{A comparison of simulation with solutions to system~\eqref{eqn:two_networks} for transmission through two overlapping static networks.  The cloud consists of simulations in a population of $1000$ individuals, with three simulations highlighted.  The other solid colored curves represent a single simulation in a population of $10000$ individuals.  The black dashed curve represets the solution to the ODE of system~\eqref{eqn:two_networks}.}
\label{fig:two_networks_soln}
\end{figure}

Figure~\ref{fig:two_networks_soln} compares simulated epidemics in a population of 10000, with two networks along which disease transmits.   In the sexual network a quarter of the population has $2$ partners, half have $3$ partners, and a quarter have $4$ partners.  In the other ``social'' network, half have $10$ partners and the other half have $20$ partners.  These numbers are not assigned independently.  The least active individuals in both networks are the same individuals.  That is  a quarter each have 2 sexual partners and 10 social partners, or $3$ sexual partners and $10$ social partners, or $3$ sexual partners and $20$ social partners, or $4$ sexual partners and $20$ sexual partners.  This yields $\psi(x,y) = (x^2y^{10} + x^3y^{10} + x^3y^{20}  +x^4y^{20})/4$.   In the first infectious phase, transmission through the sexual network happens with rate $\tau_1 = 0.5$, \ $\beta=0.1$, and $\gamma_1=1$.  Although the social network has a lower transmission rate, its much higher density means it dominates infection in this stage.  In the second stage only the sexual network transmits and $\tau_2=0.2$ and $\gamma_2=2$.   

\subsubsection{$\Ro$}
To find $\Ro$ we define $R_{so|so}$ to be the expected number of social transmissions caused given that an individual is infected through a social transmission, $R_{so|se}$ to be the expected number of social transmissions caused given that an individual is infected through a sexual transmission, and similarly $R_{se|se}$ and $R_{se|so}$.  For notational simplicity, we define $T_{so} = \beta/(\beta+\gamma_1)$ to be the probability a social contact will transmit before the infected individual moves into the second infectious stage.  We also define $T_{se} = 1 - \frac{\gamma_1}{\tau_1+\gamma_1}\frac{\gamma_2}{\tau_2+\gamma_2} $ to be the probability a sexual contact will transmit in (at least) one of the two stages.  We have
\begin{align*}
R_{se|se} &= T_{se} \frac{\psi_{xx}(1,1)}{\ave{K_{se}}}\\
R_{se|so} &= T_{se} \frac{\psi_{xy}(1,1)}{\ave{K_{so}}}\\
R_{so|se} &= T_{so} \frac{\psi_{xy}(1,1)}{\ave{K_{se}}}\\ 
R_{so|so} &= T_{so}\frac{\psi_{yy}(1,1)}{\ave{K_{so}}}
\end{align*}
Note that $\psi_{xx}(1,1) = \ave{K^2_{se} - K_{se}}$, \ $\psi_{xy}(1,1) = \ave{K_{so}K_{se}}$, and $\psi_{yy}(1,1) = \ave{K^2_{so} - K_{so}}$
Taking $N_{se}(g)$ and $N_{so}(g)$ to be the number infected by each method in generation $g$ we have
\[
\begin{pmatrix}
N_{se}(g+1)\\
N_{so}(g+1)
\end{pmatrix}
= 
\begin{pmatrix}
R_{se|se} & R_{se|so}\\
R_{so|se} & R_{so|so}
\end{pmatrix}
\begin{pmatrix}
N_{se}(g)\\
N_{so}(g)
\end{pmatrix}
\]
The dominant eigenvalue of this matrix is $\Ro$.  

\subsubsection{Final size}
The final size relation is straightforward to write out.  Rather than deriving it by solving the $t\to\infty$ limit, we derive it directly by considering the consistency relations.  Given $\theta_{se}(\infty)$ and $\theta_{so}(\infty)$, the probability that a sexual partner of $u$ is not infected is $(1-\rho) \frac{\psi_x(\theta_{se}(\infty),\theta_{so}(\infty))}{\ave{K_{se}}}$.  Similarly the probability a social partner of $u$ is not infected is $(1-\rho) \frac{\psi_y(\theta_{so}(\infty),\theta_{so}(\infty))}{\ave{K_{so}}}$.  We can state our consistency relation as follows: $\theta_{se}(\infty)$ is the probability that the partner would not transmit, or it would, but it is never infected.  A similar relation holds for $\theta_{so}(\infty)$.  Translated into equations, these become
\begin{align*}
\theta_{se}(\infty) &= 1- T_{se} + T_{se} (1-\rho) \frac{\psi_x(\theta_{se}(\infty),\theta_{so}(\infty))}{\ave{K_{se}}}\\
\theta_{so}(\infty) &= 1- T_{so} + T_{so} (1-\rho) \frac{\psi_y(\theta_{so}(\infty),\theta_{so}(\infty))}{\ave{K_{so}}}
\end{align*}
We can solve this iteratively.  From the solution, we have $S(\infty) = \psi(\theta_{se}(\infty),\theta_{so}(\infty))$.

\subsection{A dynamic sexual network with mass action mixing.}
Thus far we have considered the same underlying sexual network assumptions and modified our assumptions about the other transmission process.  Now we go back to the simple mass action assumptions for the non-sexual transmission process and focus our attention on different sexual network structures.  We begin by assuming that partnerships change in time.

We assume that the duration of partnerships is exponentially distributed with mean $1/\eta$.  When a partnership ends, the two individuals immediately form new partnerships.  Conceptually, we can think of an individual that has $k$ partners as having $k$ \emph{stubs}, which are connected to stubs of other individuals (binding sites in the language of~\cite{leung2015si,leung2016dangerous}).  A partnership terminates with rate $\eta$, and the corresponding stubs immediately join with other stubs that also terminated at the same time.  

We define $\theta$ to be the probability that for a given stub, no partner associated with that stub has ever transmitted to the test individual $u$.  As before, the probability that $u$ is susceptible is $S(t) = e^{-\xi(t)} \psi(\theta(t))$, but determining $\phi_S$, $\phi_I$, and $\phi_R$ is different.  For this we follow~\cite[chapter~8]{EoNbook}.

\begin{figure}
\begin{center}
\begin{tikzpicture}
    \node[box, fill=colorS!60] (S) at (0,1) {$S$};
    \node[box, fill=colorI!60] (I1) at (3,1) {$I_1$};
    \node[box, fill=colorI2!60] (I2) at (6,1) {$I_2$};
    \node[box, fill=colorR!60] (R) at (9,1) {$R$};
    \path [->] (S) edge node {} (I1);
    \path [->, above] (I1) edge node {$\gamma_1 I_1$} (I2);
    \path [decay] (I2) --(R) node [midway, above] {$\gamma_2 I_2$} (R);

      \node[box, fill=colorS!30] (phiS) at ( 0,-3) {$\phi_S$};
      \node[box, fill=colorI!30] (phiI1) at ( 3,-3) {$\phi_{I,1}$};
      \node[box, fill=colorI2!30] (phiI2) at ( 6,-3) {$\phi_{I,2}$};
      \node[box, fill=colorR!30] (phiR) at ( 9,-3) {$\phi_R$};
      \coordinate (break1) at (8,-1.2);
      \coordinate (break2) at (6.5,-1);
      \coordinate (break3) at (5,-1);
      \coordinate (reform1) at (4,-1);
      \coordinate (reform2) at (2.5,-1);
      \coordinate (reform3) at (1,-1.2);
      \coordinate (postreform) at (-1.6,-1.2);
      \begin{scope}[on background layer]
        \node[bigbox = {$\theta$}, fit=(phiS)(phiR)(break3)(reform1), fill=colorS!15] (Theta) {};
      \end{scope}
      \node[box,fill=colorI!60] (OmTheta) at (4.5,-5) {$1-\theta$};
      \path [->, below] (phiS) edge node {$\phi_S\frac{\psi''(\theta)}{\psi'(\theta)}\sum \tau_j \phi_{I,j} $} (phiI1);
      \path [->, below] (phiI1) edge node [midway] {$\gamma_1 \phi_{I,1}$} (phiI2);
      \path [decay] (phiI2)--(phiR) node [midway, below] {$\gamma_2 \phi_{I,2}$};
      \path [->, below, sloped] (phiI1) edge node [pos=0.5] {$\tau_1 \phi_{I,1}$}    (OmTheta);
      \path [->, below, sloped] (phiI2) edge node [pos=0.5] {$\tau_2 \phi_{I,2}$}    (OmTheta);
      \path [->, above] (break3) edge node {$\eta \theta$} (reform1);
      \path [->, above, out = 20, in = 0, sloped] (phiS) edge node [pos=0.05] {$\eta \phi_S$} (break3);
      \path [->, above, in = 0, sloped] (phiI1) edge node [pos=0.08]  {$\eta \phi_{I,1}$} (break2);
      \path [->, above, in = 340, sloped] (phiI2) edge node [pos=0.08]  {$\eta \phi_{I,1}$} (break1);
      \path [->, above, in = 340, sloped] (phiR) edge node [pos=0.3] {$\eta \phi_R$} (break1);
      \path [->, above, out=160, in = 0] (break1) edge node {} (break2);
      \path [->, above, out=180, in = 0] (break2) edge node {} (break3);
      \path [->, above, out=180, in = 0] (reform1) edge node {} (reform2);
      \path [->, above, out=180, in = 20] (reform2) edge node {} (reform3);
      \path [->, above, out = 200, sloped] (reform3) edge node [pos=0.7] {$\eta \theta \pi_S$} (phiS);
      \path [->, above, out = 200, sloped] (reform3) edge node [pos=0.95] {$\eta \theta \pi_{I,1}$} (phiI1);
      \path [->, above, out = 180, sloped] (reform2) edge node [pos=0.95] {$\eta \theta \pi_{I,2}$} (phiI2);
      \path [->, above, out = 180, in = 160, sloped] (reform1) edge node [pos=0.95] {$\eta \theta \pi_R$} (phiR);

      \node[box, fill=colorS!60] (piS) at ( -2,-6.5) {$\pi_S$};
      \node[box, fill=colorI!60] (piI1) at ( 1,-6.5) {$\pi_{I,1}$};
      \node[box, fill=colorI2!60] (piI2) at ( 4,-6.5) {$\pi_{I,2}$};
      \node[box, fill=colorR!60] (piR) at ( 7,-6.5) {$\pi_R$};
      \path [->, above] (piS) edge node {} (piI1);
      \path [->, above] (piI1) edge node {$\gamma_1\pi_{I,2}$} (piI2);
      \path [decay] (piI2) --(piR) node [midway, above] {$\gamma_2 \pi_{I,2}$};

    \node [box, fill = colorI!60] (xi) at (11, -6.5) {$\xi$};
    \node (artificial) at (9,-6.5) {};
    \path [->, above] (artificial) edge node {$\beta I_1$}   (xi);

  \end{tikzpicture}
\end{center}
\caption{Flow diagrams for mass action transmission combined with sexual network transmission.  The top diagram is as before.  The middle diagram is much as before for the network component, but because partnerships can change there are additional paths to take.  Because of this we cannot solve for $\phi_S$ explicitly and we must calculate all the fluxes in and out of $\phi_S$.  The fluxes due to finding new partners depend on the probability that the new partner has a given status.  These are tracked using the diagram in the bottom left.  The bottom right is as before.}
\end{figure}
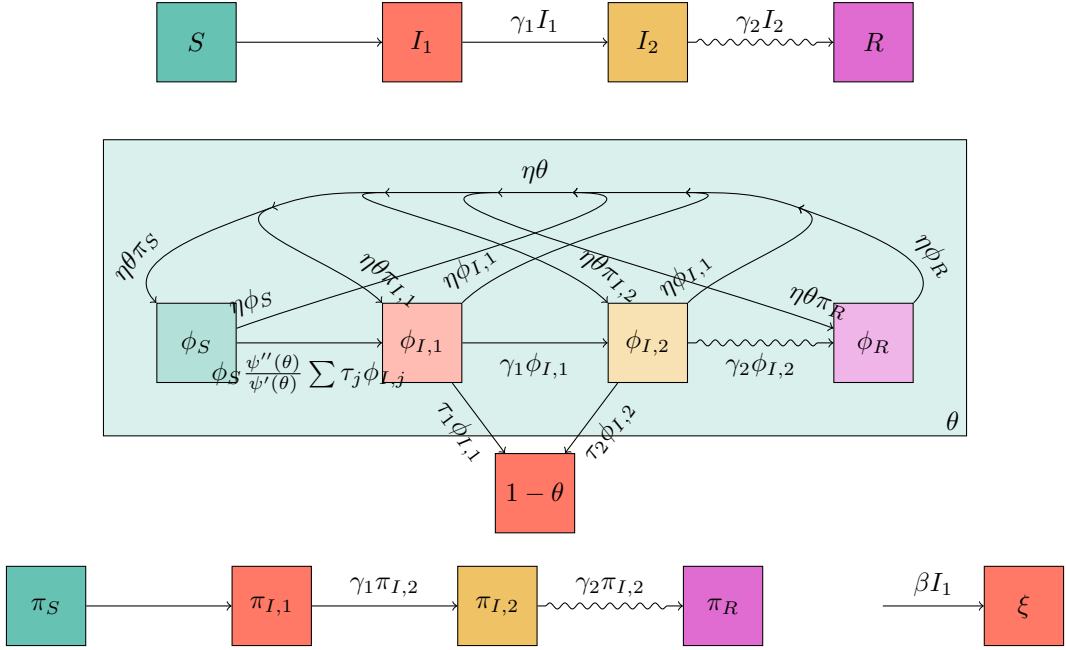

Let $v$ be a random neighbor of $u$ at time $t$.  We define $\zeta(t)$ to be the probability that the two stubs belonging to $u$ and $v$ which joins them have not carried infection to $u$ or  $v$ prior to the edge forming.  Then 
\[
\phi_S =  (1-\rho)e^{-\xi}\zeta \sum_{k_v} \frac{k_v P(k_v)}{\ave{K}} \theta^{k_v-1} = (1-\rho)e^{-\xi}\zeta \frac{\psi'(\theta)}{\ave{K}}
\]
The derivative of $\zeta$ is $\eta \theta^2 - \eta \zeta$: the rate at which such edges are created minus the rate at which such edges are destroyed by partnership turnover.

We have an option of how to build the model.  We could write $\phi_S=(1-\rho)e^{-\xi}\zeta \frac{\psi'(\theta)}{\ave{K}}$ and use a differential equation for $\zeta$.  However, we choose to eliminate $\zeta$, and instead we write
\[
\dot{\phi}_S = -\dot{\xi} \phi_S + \dot{\zeta} (1-\rho)e^{-\xi} \frac{\psi'(\theta)}{\ave{K}} + \dot{\theta}(1-\rho)e^{-\xi} \zeta \frac{\psi''(\theta)}{\ave{K}}
\]
This becomes 
\[
\dot{\phi}_S = - \dot{\xi} \phi_S + \eta \theta \pi_S - \eta \phi_S - (\tau_1\phi_{I,1} + \tau_2 \phi_{I,2})\phi_S \frac{\psi''(\theta)}{\psi'(\theta)}
\]
Ultimately, our equations are
\begin{subequations}
\label{eqn:dynamic}
\begin{align}
S &= (1-\rho) e^{-\xi} \psi(\theta)\\
I_1 &= 1- S - I_2 - R\\
\dot{I}_2 &= \gamma_1 I_1 - \gamma_2 I_2\\
\dot{R} &= \gamma_2 I_2\\
\dot{\xi} &= \beta I_1\\
\dot{\theta} &= -\tau_1 \phi_{I,1} - \tau_2 \phi_{I,2}\\
\dot{\phi}_S &= -\beta I_1 \phi_S + \eta \theta \pi_S - \eta \phi_S - (\tau_1\phi_{I,1} + \tau_2 \phi_{I,2})\phi_S \frac{\psi''(\theta)}{\psi'(\theta)}\\
\phi_{I,1} &= \theta - \phi_S - \phi_{I,2} - \phi_R\\
\dot{\phi}_{I,2} &= \gamma_1 \phi_{I,1} + \eta \theta \pi_{I,2} - (\eta + \gamma_2 + \tau_2) \phi_{I,2} \\
\dot{\phi}_{R} &= \eta \theta \pi_R + \gamma_2 \phi_{I,2} - \eta\phi_R\\ 
\pi_{S} &= (1-\rho)\frac{\theta e^{-\xi}\psi'(\theta)}{\ave{K}}\\
\pi_{I,1} &= 1 - \pi_S - \pi_{I,2} - \pi_R\\
\dot{\pi}_{I,2}  &= \gamma_1 \pi_{I,1} - \gamma_2 \pi_{I,2}\\
\dot{\pi}_R &= \gamma_2 \pi_{I,2}
\end{align}
\end{subequations}

\begin{figure}
\begin{center}
\includegraphics[width=0.5\columnwidth]{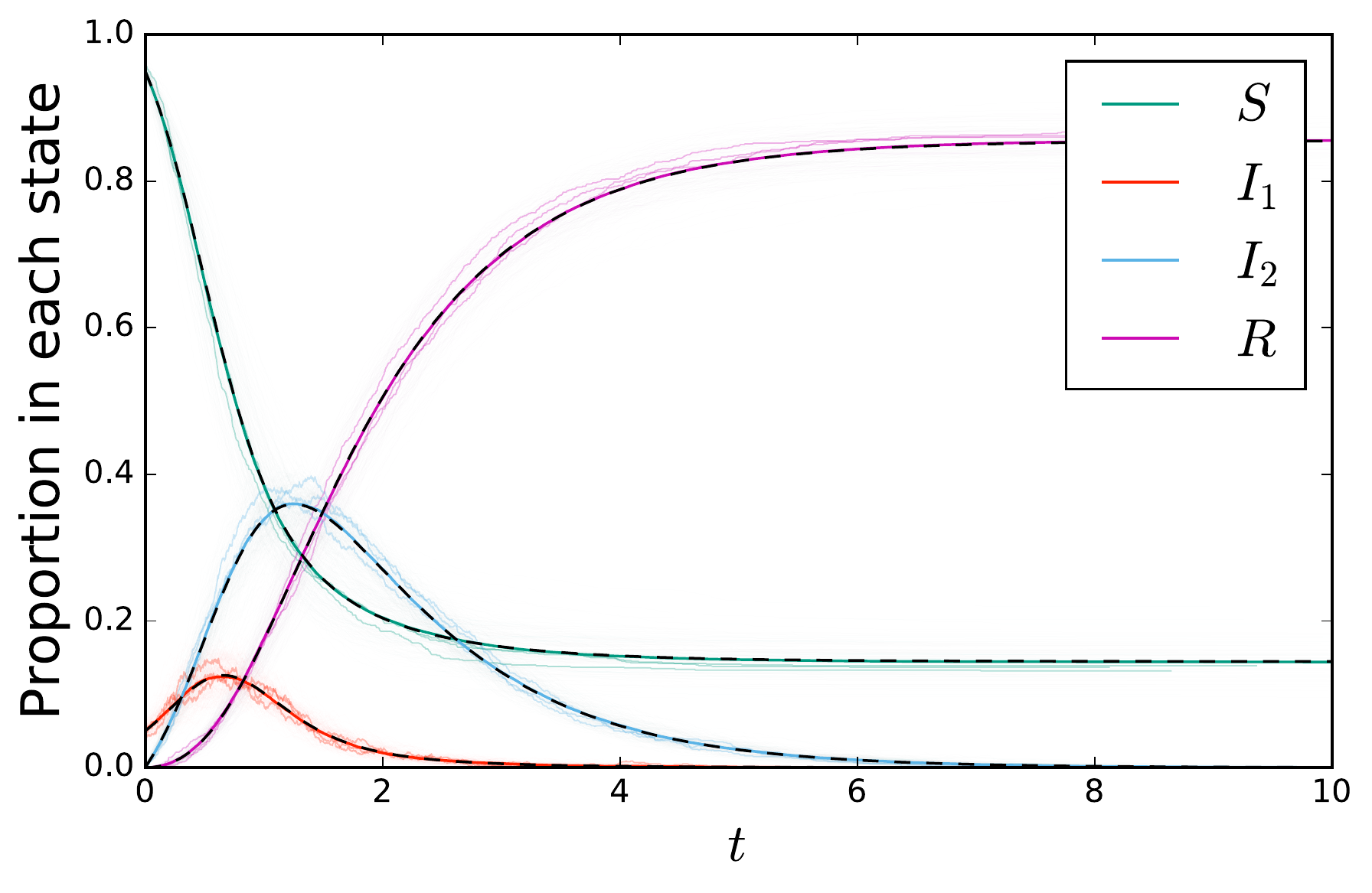}
\end{center}
\caption{A comparison of simulation with solutions to system~\eqref{eqn:dynamic} for mass action transmission combined with sexual trasmission through a dynamic network.  The cloud consists of simulations in a population of $1000$ individuals, with three simulations highlighted.  The additional solid colored curve is a single simulation in a population of $100000$ individuals.  The dashed black curve is the solution to system~\eqref{eqn:dynamic}.}
\label{fig:dynamic_soln}
\end{figure}

Figure~\ref{fig:dynamic_soln} compares simulated epidemics in a population of 10000.  Transmissions occur through a mass action mechanism and through transmission in a sexual network.  The sexual network is dynamic, with a partnership ending with rate $\eta$ and the individuals immediately finding new partners from other individuals that simultaneously ended their partnerships.  As before only the sexual network transmits after the first infectious stage.  The sexual network has $\psi(x) = (x^2+2x^3+x^4)/4$.   In the first stage, $\tau_1 = 1$, \ $\beta = 5$, and $\gamma_1=5$.   For the second stage $\tau_2=0.3$ and $\gamma_2=1$.

\subsubsection{$\Ro$}
Although $\Ro$ can be calculated for this model, the calculation is quite tedious.  We must account for the fact that when an SI partnership forms, it can transmit, the partnership can dissolve and be replaced, and retransmit again.  It become particularly tedious as we must consider the possible state of the partnership when the first transition happens and calculate different outcomes for whether it is II or SI, and take an appropriate weighted average.

For now, we note that the $\Ro$ calculated for the static network case is a lower bound.  We can find an upper bound by assuming the $\eta \to \infty$ limit such the partnerships always change prior to the next transmission.  Then when an individual becomes infected along a sexual contact we expect $\ave{K^2}/\ave{K}$ partnerships available to transmit at all times.  However when an individual becomes infected through a mass action transmission we expect $\ave{K}$ partnerships to be available.  The number of transmissions occurring for each is $\frac{\tau_1}{\gamma_1} + \frac{\tau_2}{\gamma_2}$.  

Repeating our earlier approach and counting the number of individuals infected and partnerships available to transmit, we find that the number of infected individuals in generation $g+1$ is 
\[
N_m(g+1) = \frac{\beta}{\gamma_1}N_m(g) + \left(\frac{\tau_1}{\gamma_1} + \frac{\tau_2}{\gamma_2} \right) N_{p}(g)
\]
and the number of partnerships available to transmit is
\[
N_p(g+1) = \frac{\beta}{\gamma_1}\ave{K} N_m(g) + \left(\frac{\tau_1}{\gamma_1} + \frac{\tau_2}{\gamma_2} \right) \frac{\ave{K^2}}{\ave{K}} N_p(g)
\]
Thus we have defined a matrix problem, and the dominant eigenvalue of
\[
\begin{pmatrix}
\frac{\beta}{\gamma_1} & \frac{\tau_1}{\gamma_1} + \frac{\tau_2}{\gamma_2} \\ 
\frac{\beta}{\gamma_1} \ave{K}& \left(\frac{\tau_1}{\gamma_1} + \frac{\tau_2}{\gamma_2} \right)\frac{\ave{K^2}}{\ave{K}}
\end{pmatrix}
\]
is an upper bound for $\Ro$.

\subsubsection{Final Size}
Because the partnerships are dynamic (but have finite duration) it is not possible to derive a simple final size relation like the ones we have seen before.  This is because the speed of the epidemic affects how likely a particular stub is to bring infection to a node.  If the disease spreads very quickly (with a timescale shorter than partnership duration), then the risk of transmission is tied up entirely in the likelihood that a single partner will be infected.  If that partner is infected, then perhaps many transmissions will occur (but only one will be successful).  If the disease spreads very slowly, then the infected individuals may change partners many times, and so rather than wasting many transmissions on a single individual, they transmit to many individuals.

In either extreme limit we can derive final size relations, but for any nonzero, finite partnership duration, we cannot.

\subsection{Sexual network with preferential mixing}
Finally we consider a model in which individuals select their partners preferentially according to degree.  This has been studied using pair-approximations~\cite{moreno2003epidemic}, but to our knowledge, the corresponding EBCM model has not been explicitly considered.  It can be inferred from model 2.2.3 of~\cite{miller:ebcm_structure}, and appears as an exercise in~\cite[chapter 6]{EoNbook}.  We will study the slightly more general case in which there are two infectious stages as before, and allow for the first infectious stage to also exhibit mass action transmission.

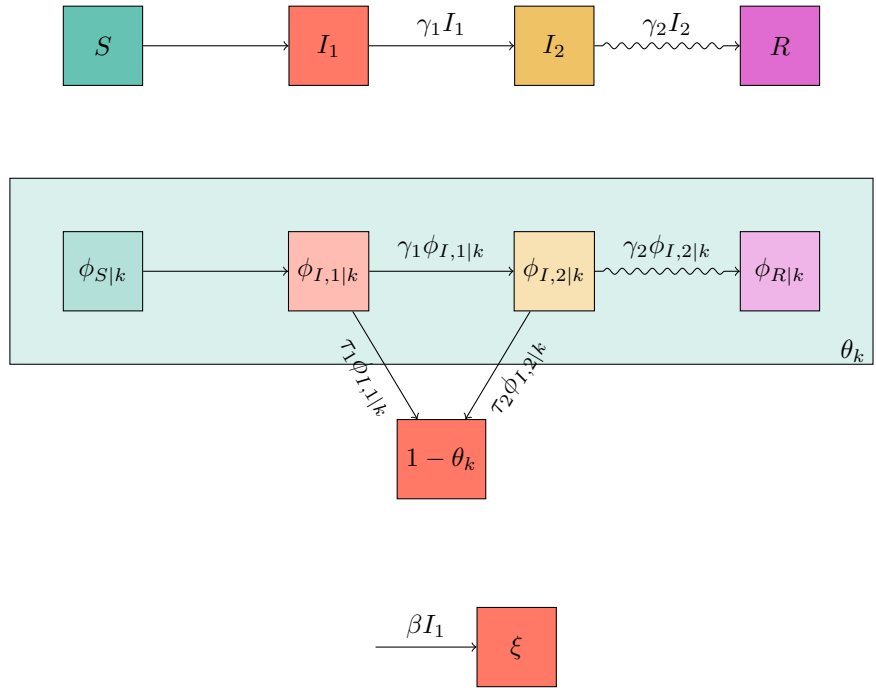
\begin{figure}
\begin{center}
\begin{tikzpicture}
    \node[box, fill=colorS!60] (S) at (0,1) {$S$};
    \node[box, fill=colorI!60] (I1) at (3,1) {$I_1$};
    \node[box, fill=colorI2!60] (I2) at (6,1) {$I_2$};
    \node[box, fill=colorR!60] (R) at (9,1) {$R$};
    \path [->] (S) edge node {} (I1);
    \path [->, above] (I1) edge node {$\gamma_1 I_1$} (I2);
    \path [decay] (I2) --(R) node [midway, above] {$\gamma_2 I_2$} (R);
    \node[box, fill=colorS!30] (phiS) at ( 0,-2) {$\phi_{S|k}$};
    \node[box, fill=colorI!30] (phiI1) at ( 3,-2) {$\phi_{I,1|k}$};
    \node[box, fill=colorI2!30] (phiI2) at ( 6,-2) {$\phi_{I,2|k}$};
    \node[box, fill=colorR!30] (phiR) at ( 9,-2) {$\phi_{R|k}$};
    \begin{scope}[on background layer]
      \node[bigbox = {$\theta_k$}, fit=(phiS)(phiR), fill=colorS!15] (Theta) {};
    \end{scope}
    \node[box,fill=colorI!60] (OmTheta) at (4.5,-4.5) {$1-\theta_k$};
    \path [->] (phiS) edge node {} (phiI1);
    \path [->, above] (phiI1) edge node {$\gamma_1 \phi_{I,1|k}$} (phiI2);
    \path [decay] (phiI2) -- (phiR) node [above, midway] {$\gamma_2\phi_{I,2|k}$};
    \path [->,below, sloped] (phiI1) edge node {$\tau_1\phi_{I,1|k}$} (OmTheta);
    \path [->,below, sloped] (phiI2) edge node {$\tau_2\phi_{I,2|k}$} (OmTheta);
    \node [box, fill = colorI!60] (xi) at (5.5, -7) {$\xi$};
    \node (artificial) at (3.5,-7) {};
    \path [->, above] (artificial) edge node {$\beta I_1$}   (xi);
  \end{tikzpicture}
\end{center}
\caption{Flow diagrams for mass action transmission combined with  sexual transmission in a preferentially mixing sexual network.  The top diagram is as before.  The middle diagram tracks $\theta_k$ so there is a corresponding equation for each degree $k$.  The bottom diagram is as before.}
\end{figure}

\begin{subequations}
\label{eqn:preferential}
\begin{align}
S &= e^{-\xi}\sum_k P(k) \theta_k^k\\
I_1 &= 1 - S - I_2 - R\\
\dot{I}_2 &= \gamma_1 I_1-\gamma_2I_2\\
\dot{R} &= \gamma_2 I_2\\
\dot{\theta}_k &= - \tau_1 \phi_{I,1|k} - \tau_2\phi_{I,2|k}\\
\phi_{S|k} &= e^{-\xi}\sum_{\hat{k}} P_n(\hat{k}|k) \theta_{\hat{k}}^{\hat{k}-1}\\
\phi_{I,1|k} &= \theta_k - \phi_{S|k} - \phi_{I,2|k} - \phi_{R|k}\\
\dot{\phi}_{I,2|k} &= \gamma_1 \phi_{I,1|k} - (\tau_2+\gamma_2)\phi_{I,2|k}\\
\dot{\phi}_{R|k} &= \gamma_2 \phi_{I,2|k}\\
\dot{\xi} &= \beta I_1
\end{align}
\end{subequations}

\begin{figure}
\begin{center}
\includegraphics[width=0.5\columnwidth]{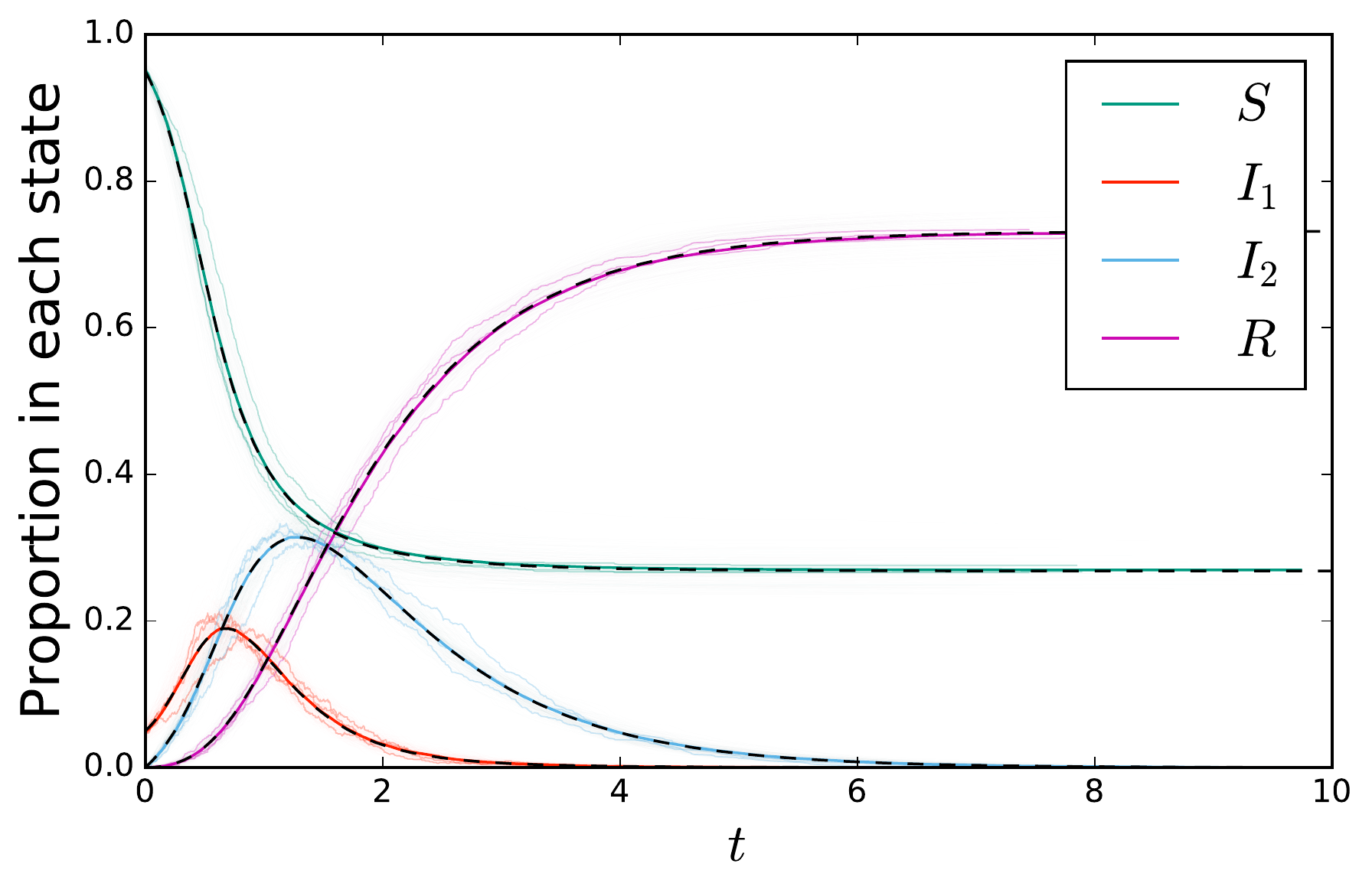}
\end{center}
\caption{A comparison of simulation with solutions to system~\eqref{eqn:preferential} for mass action transmission combined with a sexual network exhibiting preferential partnership formation.  The cloud represents $200$ simulations in a population of $1000$ individuals, with three simulations highlighted.  The additional solid colored curve comes from a simulation of $100000$ individuals.  The dashed black curve is the solution to system~\eqref{eqn:preferential}.}
\label{fig:preferential_soln}
\end{figure}

Figure~\ref{fig:preferential_soln} compares simulated epidemics in a population of 10000, with a mass action transmission mechanism and a static sexual network exhibiting preferential mixing.  The sexual network has $P(1)=P(5)=P(10)=1/3$, with the partnerships distributed so that 
\[
\begin{aligned}
&P_n(1|1) = 1/2\\
&P_n(5|1) = 3/8\\
&P_n(10|1) = 1/8
\end{aligned}
\quad
\begin{aligned}
&P_n(1|5) = 3/40\\
&P_n(5|5) = 1/2\\
&P_n(10|5) = 17/40
\end{aligned}
\quad
\begin{aligned}
&P_n(1|10)=1/80\\
&P_n(5|10)=17/80\\
&P_n(10|10)=62/80
\end{aligned}
\]
As before, the infectious period is divided into two stages.  The first infectious stage $I_1$ transmits through both mechanisms with $\tau_1=1$, \ $\beta=1$, and $\gamma_1 = 3$.  The second stage is has $\tau_2=0.1$ and $\gamma_2=1$.  

\subsubsection{$\Ro$}
To find $\Ro$ for mass action transmission combined with sexual transmission on a static preferentially mixing network, we define $N_k(g)$ to be the number of individuals having $k$ partners that are infected in generation $g$.  We again use $T = 1- \frac{\gamma_1\gamma_2}{(\tau_1+\gamma_1)(\tau_2+\gamma_2)}$ to be the probability that an infected individual transmits to a susceptible partner before recovering.  The expected number of infections  of individuals with degree  $\hat{k}$  caused through sexual transmission by an infected individual with degree $k$ is $T k P(\hat{k}|k)$.  The expected number of infections of individuals with degree $\hat{k}$ caused through mass action transmission by any infected individual is $P(\hat{k}) \beta/\gamma_1$.  Thus we conclude that
\[
N_{\hat{k}} (g+1) = \sum_k \left(T k P(\hat{k}|k) + P(\hat{k})\frac{\beta}{\gamma_1}\right) N_k(g)
\]
Writing this as a matrix equation, we have
\[
\begin{pmatrix}
N_0(g+1)\\
N_1(g+1)\\
N_2(g+1)\\
N_3(g+1)\\
\vdots\\
\end{pmatrix}
= 
\left[T\begin{pmatrix} 
0 & 0 & 0 & 0 &\cdots\\
0 & P(1|1) & 2P(1|2) & 3P(1|3)  & \cdots\\
0 & P(2|1) & 2P(2|2) & 3P(2|3)  & \cdots\\
0 & P(3|1) & 2P(3|2) & 3P(3|3) & \cdots\\
\vdots & \vdots & \vdots & \vdots & \ddots
  \end{pmatrix}
+ \frac{\beta}{\gamma_1}
\begin{pmatrix}
P(0) & P(0) & P(0) & P(0) & \cdots \\
P(1) & P(1) & P(1) & P(1) & \cdots\\
P(2) & P(2) & P(2) & P(2) & \cdots \\
P(3) & P(3) & P(3) & P(3) & \cdots \\
\vdots & \vdots & \vdots & \vdots & \ddots
\end{pmatrix}
\right]
\begin{pmatrix}
N_0(g)\\
N_1(g)\\
N_2(g)\\
N_3(g)\\
\vdots\\
\end{pmatrix}
\]
Then $\Ro$ is the dominant eigenvalue of the matrix given by the sum within the square brackets.
\subsubsection{Final Size}

Unlike the dynamic case, we can derive a final size relation for the preferential mixing model.  This is very similar to our basic combined mass action and network model.

We note that 
\[
\theta_k(\infty) = (1-T) + T \phi_{S|k}(\infty) = (1-T) + T  e^{-\xi(\infty)} \sum_{\hat{k}} P_n(\hat{k}|k) \theta_{\hat{k}}^{\hat{k}-1}
\]
and
\[
\xi(\infty) =   \frac{\beta}{\gamma_1} [1-S(\infty)] = \frac{\beta}{\gamma_1} \left[ 1- (1-\rho) e^{-\xi(\infty)} \sum_k P(k) \theta_k^k(\infty) \right]
\]
This provides the relation to solve.  Once we have this, then
\[
R(\infty) = 1- e^{-\xi(\infty)}\sum_k P(k) \theta_k^k(\infty) \, .
\]
\section{Discussion}
We have shown that it is possible to adapt the EBCM approach to consider transmission within a network combined with additional transmission outside the network.  This is particularly relevant for diseases which have a primary transmission mechanism but also a sexual transmission mechanism.  Ebola and Zika are recently emerging diseases which demonstrate the potential importance of this mechanism.  Particularly due to the fact that Zika causes its worst complications following infection of pregnant women, this is an important factor to include.

For both Zika and Ebola, the observed examples of sexual transmission would likely not have been recognized if exposure rates were higher.  So it is likely that sexual transmission may exist for other diseases as well.  When we think about elimination of a disease, or the emergence of a new disease, the potential for sexual transmission to lead to transmission long after an individual appeared to recover is a potentially important factor to incorporate in models, especially if there may be questions of whether sexual transmission is sufficient to maintain transmission~\cite{zika_sti_failure}.

We have developed several low-dimensional models which capture a wide range of potential ways that sexual transmission can interact with another transmission mechanism.  Using the techniques shown here, it is straightforward to modify the models to account for other interactions.  

\section{Acknowledgments}
This work was funded by the Global Good Fund
through the Institue for Disease Modeling and by a Larkins
Fellowship from Monash University

\bibliographystyle{plain}
\bibliography{abbreviations,NetworkEpidemics}
\end{document}